\documentclass[aps,prd,10pt,twocolumn,preprintnumbers,tightenlines,superscriptaddress,nofootinbib,showpacs,longbibliography,floatfix
]{revtex4-1}
\usepackage{amssymb,latexsym}
\usepackage{amsmath,amsbsy,bbm}
\usepackage{epsfig,bm,color}
\usepackage{graphicx,comment}
\begin{document}

%%%%%%%%%%%%%%%%%%%%%%%%%%%%%%%%%%%%%%%%%%%%%%%%%%%%%%%
\def\a{{\alpha}}
\def\b{{\beta}}
\def\d{{\delta}}
\def\D{{\Delta}}
\def\X{{\Xi}}
\def\e{{\varepsilon}}
\def\g{{\gamma}}
\def\G{{\Gamma}}
\def\k{{\kappa}}
\def\l{{\lambda}}
\def\L{{\Lambda}}
\def\m{{\mu}}
\def\n{{\nu}}
\def\o{{\omega}}
\def\O{{\Omega}}
\def\S{{\Sigma}}
\def\s{{\sigma}}
\def\th{{\theta}}

%%%%%%%%%%%%%%%%%%%%%%%%%%%%%%%%%%%%%%%%%%%%%%%%%%%%%%%
\def\cF{{\mathcal F}}
\def\cS{{\mathcal S}}
\def\cC{{\mathcal C}}
\def\cB{{\mathcal B}}
\def\cT{{\mathcal T}}
\def\cQ{{\mathcal Q}}
\def\cL{{\mathcal L}}
\def\cO{{\mathcal O}}
\def\cA{{\mathcal A}}
\def\cQ{{\mathcal Q}}
\def\cR{{\mathcal R}}
\def\cH{{\mathcal H}}
\def\cW{{\mathcal W}}
\def\cM{{\mathcal M}}
\def\cD{{\mathcal D}}
\def\cN{{\mathcal N}}
\def\cP{{\mathcal P}}
\def\cK{{\mathcal K}}
\def\cI{{\mathcal{I}}}
\def\cJ{{\mathcal{J}}}

%%%%%%%%%%%%%%%%%%%%%%%%%%%%%%%%%%%%%%%%%%%%%%%%%%%%%%%
\def\fA{{\mathfrak A}}
\def\fB{{\mathfrak B}}
\def\fC{{\mathfrak C}}

%%%%%%%%%%%%%%%%%%%%%%%%%%%%%%%%%%%%%%%%%%%%%%%%%%%%%%%
\def\ol#1{{\overline{#1}}}
\def\eqref#1{{(\ref{#1})}}

%%%%%%%%%%%%%%%%%%%%%%%%%%%%%%%%%%%%%%%%%%%%%%%%%%%%%%%
\def\Tr{\text{Tr}}
\def\Pslash{P\hskip-0.65em /}

 %%%%%%%%%%%%%%%%%%%%%%%%%%%%%%%%%%%%%%%%%%%%%%%%%%%%%%%
%%%%%%%%%%%%%%%%%%%%%%%%%%%%%%%%%%%%%%%%%%%%%%%%%%%%%%%
 \title{Violation of Positivity Bounds in Models of Generalized Parton Distributions}

\author{Brian~C.~Tiburzi}
\email[]{$\texttt{btiburzi@ccny.cuny.edu}$}
\affiliation{
Department of Physics,
        The City College of New York,
        New York, NY 10031, USA}
\affiliation{
Graduate School and University Center,
        The City University of New York,
        New York, NY 10016, USA}
        
\author{Gaurav Verma}
\email[]{$\texttt{Gaurav.Verma36@qmail.cuny.edu}$}
\affiliation{
Department of Physics, 
Queens College of The City University of New York, 
Queens, NY 11367, USA}

\date{\today}

%%%%%%%%%%%%%%%%%%%%%%%%%%%%%%%%%%%%%%%%%%%%%%%%%%%%%%%
%%%%%%%%%%%%%%%%%%%%%%%%%%%%%%%%%%%%%%%%%%%%%%%%%%%%%%%

%\pacs{}

\begin{abstract}
As with parton distributions, 
flexible phenomenological parameterizations of generalized parton distributions (GPDs) are essential for their extraction from data. 
The large number of constraints imposed on GPDs make simple Lorentz covariant models viable;
but, 
such models are often incomplete in that they employ the impulse approximation. 
Using the GPD of the pion as a test case, 
we show that the impulse approximation can lead to violation of the positivity bound required of GPDs.  
We focus on a particular model of the pion bound-state vertex that was recently proposed
and demonstrate that satisfying the bound is not guaranteed by Lorentz covariance. 
Violation of the positivity bound is tied to a problematic mismatch between the behavior of the quark distribution at the endpoint
and the crossover value of the GPD. 
\end{abstract}
\maketitle
 
%%%%%%%%%%%%%%%%%%%%%%%%%%%%%%%%%%%%%%%%%%%%%%%%%%%%%%%
%%%%%%%%%%%%%%%%%%%%%%%%%%%%%%%%%%%%%%%%%%%%%%%%%%%%%%%
%%%%%%%%%%%%%%%%%%%%%%%%%%%%%%%%%%%%%%%%%%%%%%%%%%%%%%%
%%%%%%%%%%%%%%%%%%%%%%%%%%%%%%%%%%%%%%%%%%%%%%%%%%%%%%%

%%%%%%%%%%%%%%%%%%%%%%%%%%%
\section{Introduction}                                                 %
%%%%%%%%%%%%%%%%%%%%%%%%%%%

Generalized parton distributions%
~\cite{Mueller:1998fv,Ji:1996ek,Ji:1996nm,Radyushkin:1996nd,Radyushkin:1996ru} 
contain the physics of form factors and parton distributions, 
and thereby allow for the study of correlations between transverse position and longitudinal momentum inside hadrons. 
In impact-parameter space, 
GPDs elegantly describe the transverse structure of fast moving hadrons%
~\cite{Burkardt:2000za,Burkardt:2002hr,Diehl:2002he}. 
Intense activity has been generated in this field,
which is largely due to the ability to measure GPDs in deeply virtual Compton scattering, 
and resolve the angular momentum content of hadrons. 
A number of insightful reviews have appeared on the subject, 
see, 
for example,
Refs.~\cite{Goeke:2001tz,Diehl:2003ny,Belitsky:2005qn,Boffi:2007yc}.

From a theoretical perspective, 
GPDs are rather complicated objects to model. 
As with ordinary parton distribution functions, 
flexible phenomenological parameterizations would be welcome to aid in their extraction from data. 
The large number of constraints imposed on GPDs, 
however, 
makes such parameterizations challenging to devise. 
One such non-trivial constraint required of GPDs is a general bound due to the positivity of the norm on the hadronic light-front Fock space. 
These so-called positivity bounds are introduced and discussed in several works%
~\cite{Martin:1997wy,Ji:1998pc,Pire:1998nw,Radyushkin:1998es,Pobylitsa:2001nt,Pobylitsa:2002gw}.

As a simple example of a positivity bound, 
consider the mixed coordinate-space density in a quantum mechanical state 
$| \psi \rangle$, 
which has the form
$\rho(\bm{r}', \bm{r}) = \langle \bm{r}' | \psi \rangle \langle \psi | \bm{r} \rangle$. 
The corresponding density is strictly positive semidefinite, 
$\rho(\bm{r}) \equiv |\langle \bm{r} | \psi \rangle|^2 \geq 0$. 
Accordingly, 
the mixed density satisfies the bound
\begin{equation}
|\rho(\bm{r}', \bm{r}) |
\leq
\sqrt{\rho(\bm{r}') \rho(\bm{r})}
,\end{equation}
as a consequence of the Schwarz inequality. 
In the case of GPDs, 
the underlying parton distribution, 
$q(x)$,
is a probability distribution and thereby satisfies the positivity condition
$q(x) \geq 0$. 
The GPD is similar in structure to a mixed plus-momentum density in the bound state. 
With symmetrical kinematic variables
and for 
$x>\xi$, 
the incoming parton has momentum fraction 
$x_i = (x + \xi) / (1 + \xi)$,
and the outgoing parton has momentum fraction 
$x_f = (x- \xi) / (1 - \xi)$. 
On account of the Cauchy-Schwarz inequality, 
the pion GPD, 
which we denote by
$H(x,\xi,t)$,
is bounded by
\begin{equation}
|H(x,\xi,t)| 
\leq 
\sqrt{  q ( x_f ) q ( x_i )}
,\end{equation}
when 
$x > \xi$. 
If the GPD is expressed in terms of quark light-front wavefunctions, 
this positivity bound is an immediate consequence that arises from the convolution of diagonal Fock-state wavefunctions%
~\cite{Brodsky:2000xy,Diehl:2000xz}. 
While not manifestly Lorentz covariant, 
the light-front Fock-space expansion makes the positivity bound transparent. 
On the other hand, 
the large number of constraints on GPDs stemming from Lorentz invariance are best satisfied within Lorentz covariant frameworks, 
for which purpose the double distributions (DDs) have been devised%
~\cite{Radyushkin:1997ki,Radyushkin:1998es}. 
Such frameworks, 
however, 
obfuscate the positivity bound. 
This situation is addressed in the present work;
and, 
we argue that the impulse approximation, 
while Lorentz covariant, 
is insufficient to describe the higher Fock components self consistently.
As a result, 
the positivity bound on GPDs can be violated in the impulse approximation, 
and we use the particular model of 
Ref.~\cite{Fanelli:2016aqc} 
to exemplify this fact.

The calculation of model GPDs in Lorentz covariant frameworks is not entirely new,
however, 
the line of investigation pursued in 
Ref.~\cite{Fanelli:2016aqc} 
represents a physically motivated departure from earlier models. 
Such models have largely employed the point-like bound-state vertex arising in the Nambu--Jona-Lasinio model, 
a couple examples of which are 
Refs.~\cite{Theussl:2002xp,Broniowski:2007si}. 
While simple, 
the point-like Ansatz provides a solution of the Bethe-Salpeter equation with a contact interaction that includes a binding effect. 
As the particular contact interaction cannot be gauged, 
the point-like model is complete in the impulse approximation. 
The corresponding model GPDs should then satisfy all known constraints, 
modulo difficulties arising from the regularization of ultraviolet divergences. 
The potential danger of these subtractions on the positivity bound was discussed in 
Ref.~\cite{Broniowski:2007si}, 
where it was found that a particular Pauli-Villars subtraction%
~\cite{Davidson:1994uv}
maintains the positivity bound on GPDs.
As pointed out in
Ref.~\cite{Pobylitsa:2002vw},  
one can construct further consistent models of GPDs by summing over these basic contributions evaluated at different constituent masses.  
This is the essence of the consistency of the spectral quark model GPDs computed in 
Ref.~\cite{Broniowski:2007si};
however, 
it must be stressed that, 
unlike two-point functions,  
GPDs do not possess a spectral representation. 
In contrast to these models, 
that employed in
Ref.~\cite{Fanelli:2016aqc}
introduces a 
\emph{non} 
point-like Bethe-Salpeter vertex for the bound state. 
This covariant smearing has the salient feature that ultraviolet divergences are cured; 
however, 
such an Ansatz arises from a non-trivial kernel. 
Without knowledge of the underlying dynamics, 
one does not know how to gauge the kernel and such contributions must be omitted. 
Due to this malady, 
the model is incomplete in the impulse approximation. 
Consequently the positivity bound cannot be guaranteed; 
and, 
indeed, 
we find that it is (rather severely) violated.%
\footnote{ 
Beyond writing down flexible parameterizations of GPDs, 
there is also interest in computing GPDs using dynamical models. 
Considerable interest has been recently generated%
~\cite{Karmanov:2005nv,Carbonell:2008tz,Carbonell:2017kqa}
by the ability to obtain light-front solutions from Bethe-Salpeter wavefunctions by way of the Nakanishi integral representation%
~\cite{Nakanishi:1963zz,Nakanishi:1969ph}.
In light of the present work, 
the positivity bound on GPDs provides an essential test of the consistency of form factors and quark distributions obtained in such an approach. 
In another direction, 
valence quark distributions of pseudoscalar mesons have been obtained from solutions to a model based on truncated Dyson-Schwinger equations%
~\cite{Nguyen:2011jy}. 
While incorporating some of the features of QCD, 
it is already recognized that the impulse approximation is insufficient to describe GPDs within this approach~\cite{Mezrag:2014jka}.
}

The organization of this work is as follows. 
We begin with a short reminder about the definitions of the Lorentz invariant 
DDs in 
Sec.~\ref{s:DD}, 
and determine the 
DDs for the covariant model of the pion proposed in 
Ref.~\cite{Fanelli:2016aqc}. 
Technical details in the computation of the DDs are relegated to Appendix~\ref{s:A}. 
The DDs are used to compute the pion GPD, 
and the positivity bound is shown to be violated, 
both in strong and weak forms. 
Investigating the cause of this violation leads us to the light-front 
Fock-space representation of 
GPDs in Sec.~\ref{s:LFWF}. 
Here, 
an analysis of the pole structure leading to the light-front representation of the GPD is given. 
We argue that positivity violation arises from the mismatch between the endpoint behavior of the quark distribution and the crossover value of the GPD. 
Appendix~\ref{s:B} provides details about light-front spinors, normalization factors, and other conventions employed. 
A brief summary, 
which is given in Sec.~\ref{s:summy},
concludes this work.

%%%%%%%%%%%%%%%%%%%%%%%%%%%%%%%%%%%%%%%
\section{The Model and its Double Distributions}
\label{s:DD}
%%%%%%%%%%%%%%%%%%%%%%%%%%%%%%%%%%%%%%%

While GPDs are not Lorentz invariant objects, 
they nonetheless inherit a number of constraints related to the underlying Lorentz covariance of their defining QCD matrix elements. 
These constraints 
(in particular, the polynomiality of GPD moments) 
are elegantly satisfied by employing the DD representation. 
The DD for a given model is Lorentz invariant, 
and the GPD is obtained by non-covariant integration over a slice in the space where the DD has support%
~\cite{Radyushkin:1997ki,Radyushkin:1998es}.

As our focus is with Lorentz covariant models for GPDs, 
we begin with the DD representation. 
Pertinent definitions and conventions for pion DDs are reviewed, 
followed by the determination of the DDs in a covariant model of the poin. 
The positivity bound is then tested in this model, 
and found to be violated.

%%%%%%%%%%%%%%%%%%%%%%%%%%%%%%%%%%%%%%%
\subsection{Double Distributions}

Matrix elements of twist-two operators with a $t$-channel momentum transfer define moments
which can be summed into the so-called DDs. 
Using the symmetric derivative, 
which we define by
$\overset{\leftrightarrow}{D} {}^\mu = \frac{1}{2} \left( \overset{\rightarrow}{D} {}^\mu - \overset{\leftarrow}{D} {}^\mu\right)$, 
the quark bilinear twist-two operators have the form
\begin{equation}
\mathcal{O}^{\mu \mu_1 \cdots \mu_n}
= 
\ol \psi (0) \gamma^{\big\{ \mu} i \overset{\leftrightarrow}{D} {}^{\mu_1} \cdots i \overset{\leftrightarrow}{D} {}^{\mu_n \big\}} \psi(0)
\label{eq:twistO}
,\end{equation} 
where the curly brackets denote the complete symmetrization and trace subtraction of the enclosed Lorentz indices. 
Matrix elements, 
$\cM^{\mu \mu_1 \cdots \mu_n}$,
of these operators within the pion are defined by
\begin{eqnarray}
\cM^{\mu \mu_1 \cdots \mu_n}
\equiv
\Big\langle P'  \Big|
\mathcal{O}^{\mu \mu_1 \cdots \mu_n}
\Big| P \Big\rangle
,\end{eqnarray}
where 
$P' = P + \Delta$;
and,
the matrix elements can be parameterized in the form
\begin{eqnarray}
\cM^{\mu \mu_1 \cdots \mu_n}
&=&
\sum_{k = 0}^n 
\begin{pmatrix} n \\ k \end{pmatrix}
\left( 2 \ol P A_{nk} - \Delta B_{nk} \right)^{\big\{ \mu}
\ol P {}^{\mu_1} \cdots \ol P {}^{\mu_{n-k}}
\notag 
\\
&& \phantom{sp} \times 
\left(-\frac{\Delta}{2}\right)^{\mu_{n-k+1}}
\cdots
\left(-\frac{\Delta}{2}\right)^{\mu_{n} \big\}}
,\label{eq:DDecomp}
\end{eqnarray}
where the $(nk)$-th moments are Lorentz invariant functions of the $t$-channel momentum transfer, 
$A_{nk} = A_{nk}(t)$
and
$B_{nk} = B_{nk}(t)$, 
with 
$t = \Delta^2$. 
The momentum 
$\ol P_\mu$
is defined to be the average between the initial and final states, 
$\ol P_\mu = \frac{1}{2} ( P' + P)_\mu$. 
For 
$n = k = 0$, 
the matrix element is simply that of the vector current. 
Consequently, 
we have
$A_{00}(t) = F(t)$, 
where
$F(t)$
is the vector form-factor of the pion, 
and
$B_{00}(t) = 0$, 
due to vector-current conservation. 
Time-reversal invariance
restricts the allowed values of 
$k$ 
in the binomial sums. 
For the 
$A_{nk}(t)$
moments, 
$k$ must be even;
while, 
for the 
$B_{nk}(t)$ 
moments, 
$k$ 
must be odd.

The DDs are generating functions for the moments of twist-two operators. 
Written in terms of the moments, 
we can define two\footnote{
As it stands, 
there is freedom in the decomposition of moments,  
Eq.~\eqref{eq:DDecomp}, 
which ultimately implies that the DDs
$F(\b,\a,t)$
and
$G(\b,\a,t)$
are not unique. 
This redundancy is akin to gauge freedom%
~\cite{Teryaev:2001qm}, 
and a minimal gauge is one in which there is only an 
$F$-type DD,
and what is called the 
$D$-term%
~\cite{Polyakov:1999gs}, 
which reduces the $G$-type DD to a 
$\delta(\b)$ 
contribution. 
There is a way, furthermore, 
to write these two contributions as the projection of a single function, 
see
Ref.~\cite{Belitsky:2000vk}.
Nonetheless, 
we use two DDs for computational ease, 
and note that the more minimal descriptions can be straightforwardly obtained therefrom.  
}
DDs
$F(\b,\a,t)$
and
$G(\b,\a,t)$
as
\begin{eqnarray}
A_{nk}(t)
&=&
\int_0^\beta d\b
\int_{-1 + \b}^{+1 - \b}
d\a
\,
\b^{n-k} \a^{k}
\,
F(\b, \a, t),
\notag \\
B_{nk}(t)
&=&
\int_0^\beta d\b
\int_{-1 + \b}^{+1 - \b}
d\a
\,
\b^{n-k} \a^{k}
\,
G(\b, \a, t)
\label{eq:FG}
,\end{eqnarray}
which satisfy the 
$\a$-symmetry properties
$F(\b, \a, t) = F(\b, -\a, t)$
and
$G(\b, \a, t) =  - G(\b, -\a, t)$. 
These properties express the consequences of time-reversal invariance. 
Throughout, 
we consider 
$\beta >0$
for the quark distribution in the pion.

Using a vector $z_\mu$ that is light-like, 
$z^2 = 0$, 
and such that 
$z_\mu = \frac{\lambda}{\sqrt{2}}  (1, 0, 0, 1)$,  
we can sum the moments into the matrix element of the quark bilocal operator%
\footnote{
With QCD gauge interactions, 
one needs to assume the light-cone gauge,  
$z \cdot A = 0$, 
to arrive at Eq.~\eqref{eq:bilocal0}, 
otherwise a gauge link proportional to
$z \cdot U \left(-\frac{z}{2},\frac{z}{2} \right)$
will appear in the bilocal operator in order to maintain gauge invariance.  
}
\begin{eqnarray}
\mathcal{O}(z)
&=&
\ol \psi \left(- \frac{z}{2} \right) \rlap\slash  z \, \psi \left( \frac{z}{2} \right)
\notag \\
&=&
\sum_{n=0}^\infty
\frac{(-i)^n}{n!}
z_\mu z_{\mu_1} \cdots z_{\mu_n} \mathcal{O}^{\mu \mu_1 \cdots \mu_n}
\label{eq:bilocal0}
,\end{eqnarray}
which, 
on account of 
Eq.~\eqref{eq:DDecomp},  
can be written in terms of the generating functions in the form
\begin{eqnarray}
\Big\langle  P' \Big|
\mathcal{O}(z)
\Big| P \Big\rangle
=
\int_0^1 d\b \int_{-1 + \b}^{+1 - \b} d\a
\, 
e^{ - i \left( \b \ol P - \a \frac{\Delta}{2} \right) \cdot z}
\notag\\
\times
\Big[
2 \ol P \cdot z
\, 
F(\b, \a, t)
- 
\Delta \cdot z 
\, 
G(\b, \a, t)
\Big]
.\qquad
\end{eqnarray}
The pion GPD is defined from the Fourier transform of this light-cone correlation of quark fields. 
Writing the skewness variable as
\begin{equation}
\xi = - \frac{1}{2} \Delta \cdot z  / \, \ol P \cdot z
,\end{equation} 
we have the conventional definition of the
GPD as an off-forward matrix element of the bilocal operator
\begin{equation}
H(x, \xi, t)
\equiv
\int \frac{d\lambda}{4 \pi}
e^{ i x \ol P \cdot z}
\Big\langle P' \Big|
\mathcal{O}(z)
\Big| P  \Big\rangle
\label{eq:bilocal}
.\end{equation}
Expressing the GPD in terms of its underlying Lorentz invariant DDs, 
we have the relation
\begin{eqnarray}
H(x,\xi,t)
&=&
\int_0^1 d\b \int_{-1 + \b}^{+1 - \b} d \a
\, \delta( x - \b - \xi \a)
\notag \\
&& \phantom{space} \times \Big[ F(\b, \a, t) + \xi \, G(\b, \a, t) \Big]
\label{eq:GPDD}
.\end{eqnarray}

All constraints on the GPDs associated with Lorentz invariance are built into the DD representation that appears in 
Eq.~\eqref{eq:GPDD}. 
For example, 
notice that the 
$n$-th moment of the GPD with respect to $x$
\begin{eqnarray}
\int_0^1 dx \, x^n H(x, \xi, t)
&=&
\int_0^1 d\b \int_{-1 + \b}^{+1 - \b} d\a  \, (\b + \xi \a)^n 
\notag \\
&&  \times \Big[ F(\b, \a, t) + \xi \, G(\b, \a, t) \Big]
, \quad \end{eqnarray} 
is at most an $n$-th [$(n+1)$-st] degree polynomial in 
$\xi$, 
for 
$n$ even [odd]. 
The zeroth moment, 
which must be 
$\xi$
independent,  
produces the so-called GPD sum rule for the form factor
\begin{equation}
\int_0^1 dx \, H(x, \xi, t)
=
F(t)
,\end{equation}
having used 
$B_{00}(t) = 0$.
Finally, 
the quark distribution function, 
which we denote by 
$q(x)$,
is contained in the forward limit, 
$\Delta_\mu = 0$,  
of the GPD
\begin{equation}
q(x) 
=
H(x,0,0)
=
\int_{-1 + x}^{+1 - x} d\a 
\, 
F(x,\a,0)
\label{eq:q}
.\end{equation}
Notice that contributions from 
$G(\b,\a,t)$
to both the form factor and quark distribution vanish due to time-reversal invariance. 
In this way, 
GPDs naturally contain additional information unconstrained by their zeroth moment and forward limit.%
\footnote{
It was pointed out in 
Ref.~\cite{Tiburzi:2004qr}
that the ambiguity inherent in defining DDs can be used in reverse to reduce the entirety of the
$F$-type DD to a contribution proportional to 
$\delta(\a)$. 
In this representation, 
all of the skewness dependence of the GPD arises from the 
$G$-type DD, 
specifically in the form
$
H(x, \xi,t) = F(x,0,t) + \xi \int_0^1 d\b \int_{-1 + \b}^{+1 - \b} d\a \, \delta ( x - \beta - \xi \a) \, \tilde{G}(\b,\a,t)
$. 
The only constraint on the skewness dependence is through the time-reversal condition, 
$\tilde{G}(\b,\a,t) = - \tilde{G}(\b,-\a,t)$;
and, 
of course, 
the positivity bound. 
} 
Consequently, 
the positivity bound provides essential guidance for constructing consistent models for GPDs.

%%%%%%%%%%%%%%%%%%%%%%%%%%%%%%%%%%%%%%%
\subsection{The Model \& Violation of the Positivity Bound}

The model of the pion proposed in 
Ref.~\cite{Fanelli:2016aqc}, 
employs an Ansatz for the covariant Bethe-Salpeter wavefunction. 
It is written in the form 
\begin{equation}
\Psi(k,P)
=
S(k) i \gamma_5 \Gamma(k,P) S(k-P)
\label{eq:model}
,\end{equation}
where
$S(k)$
denotes the constituent quark propagator, 
which is assumed to have the free-particle form
\begin{equation}
S(k) = \frac{i}{\rlap \slash k - m + i \e}
,\end{equation}
and
$\Gamma(k,P)$
is the Bethe-Salpeter vertex. 
Within this particular model, 
the latter is taken as 
\begin{equation}
\Gamma(k,P)
\equiv
\cN \,
[k^2 - m_R^2 + i \e]^{-1}
[(k-P)^2 - m_R^2 + i \e]^{-1}
\label{eq:BSvertex}
,\end{equation}
where 
$\cN$
is a normalization factor.
The covariant vertext is symmetric under the interchange of quark momentum, 
$\G(P-k,P) = \G(k,P)$, 
as one expects for isospin symmetric valence quarks. 
The mass 
$m_R$
is a model parameter,  
which additionally serves as a built-in regulator for ultraviolet divergences.%
\footnote{
Notice that because this is an ingredient of the model, 
one does not attempt to take the limit
$m_R \to \infty$
to produce regularization independent results. 
The covariant smearing of the ultraviolet behavior of the vertex is assumed to be due to the physical structure of the bound state. 
} 
Previous covariant models, 
see,
for example,  
Ref.~\cite{Frederico:1992ye},
have adopted a point-like vertex between the quarks and pion.
While there is a binding effect in such Nambu--Jona-Lasinio type models, 
the point-like bound-state vertex is an approximation that the model of Ref.~\cite{Fanelli:2016aqc} seeks to remedy. 
Notice that beyond 
$\gamma_5$, 
there are additional structures that can contribute to the Bethe-Salpeter vertex, 
but these have been dropped in the interest of simplicity.

%%%%%%%%%%%%%%%%%%%%%%%%%%%%%%%%%%%%%%%
%
%
%
\begin{figure}
%%%%%%%%%%%%
%%%%%%%%%%%%
\resizebox{0.6\linewidth}{!}{
%%%%%%%%%%%%
%%%%%%%%%%%%
\includegraphics{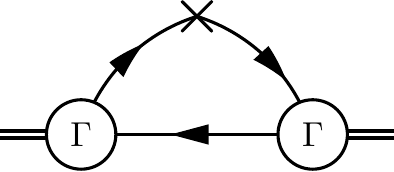}
}
\caption{
Required Feynman diagram for bound-state matrix elements of twist-two operators computed within the impulse approximation. 
Pions are depicted by double lines, with the bound-state vertices labelled by 
$\Gamma$. 
Quarks are depicted by directed lines, with insertion of the quark bilocal operator appearing in 
Eq.~\eqref{eq:twistO} 
shown by a cross.}
\label{f:triangle}
\end{figure}
%
%
%
%%%%%%%%%%%%%%%%%%%%%%%%%%%%%%%%%%%%%%%

The DDs, 
and in turn the GPDs, 
can be computed from matrix elements of the twist-two operators, 
Eq.~\eqref{eq:DDecomp}. 
Because the model employs an Ansatz for the Bethe-Salpeter vertex, 
the DDs are computed within the impulse approximation. 
The triangle diagram formed from the insertion of the twist-two operators is shown in Fig.~\ref{f:triangle}. 
Using the Bethe-Salpeter wavefunction, 
these matrix elements are given by
\begin{eqnarray}
\cM^{\mu \mu_1 \cdots \mu_n}
&=&
\int_k
\,
\Tr \Big[ \Psi(k,P) S^{-1}(k-P) \ol \Psi(k', P') \gamma^{\big\{ \mu} \Big]
\notag \\
&& 
\times 
(k + \D / 2)^{\mu_1} \cdots (k + \D / 2)^{\mu_n \big\}}
\label{eq:mel}
,\end{eqnarray}
where the shorthand 
$\int_k$
indicates the Minkowski four-momentum integration, 
$\int_k = \int d^4k / (2 \pi)^4$. 
The primed variables are defined to be boosted by the momentum transfer
$\D$, 
so that 
$k' = k + \D$ 
for the struck quark, 
along with
$P' = P + \D$
for the final-state pion. 
With a real-valued normalization factor, 
$\cN \in \mathbb{R}$, 
the conjugate Bethe-Salpeter wavefunction is given by
\begin{equation}
\ol \Psi (k',P') 
= 
S(k'-P') i  \gamma_5 \Gamma(k',P') S(k')
.\end{equation} 
Notice the equality of the momentum differences for the spectator quark, 
$k'-P' = k-P$.

%%%%%%%%%%%%%%%%%%%%%%%%%%%%%%%%%%%%%%%
%
%
%
\begin{figure}
%%%%%%%%%%%%
%%%%%%%%%%%%
\resizebox{\linewidth}{!}{
%%%%%%%%%%%%
%%%%%%%%%%%%
\includegraphics{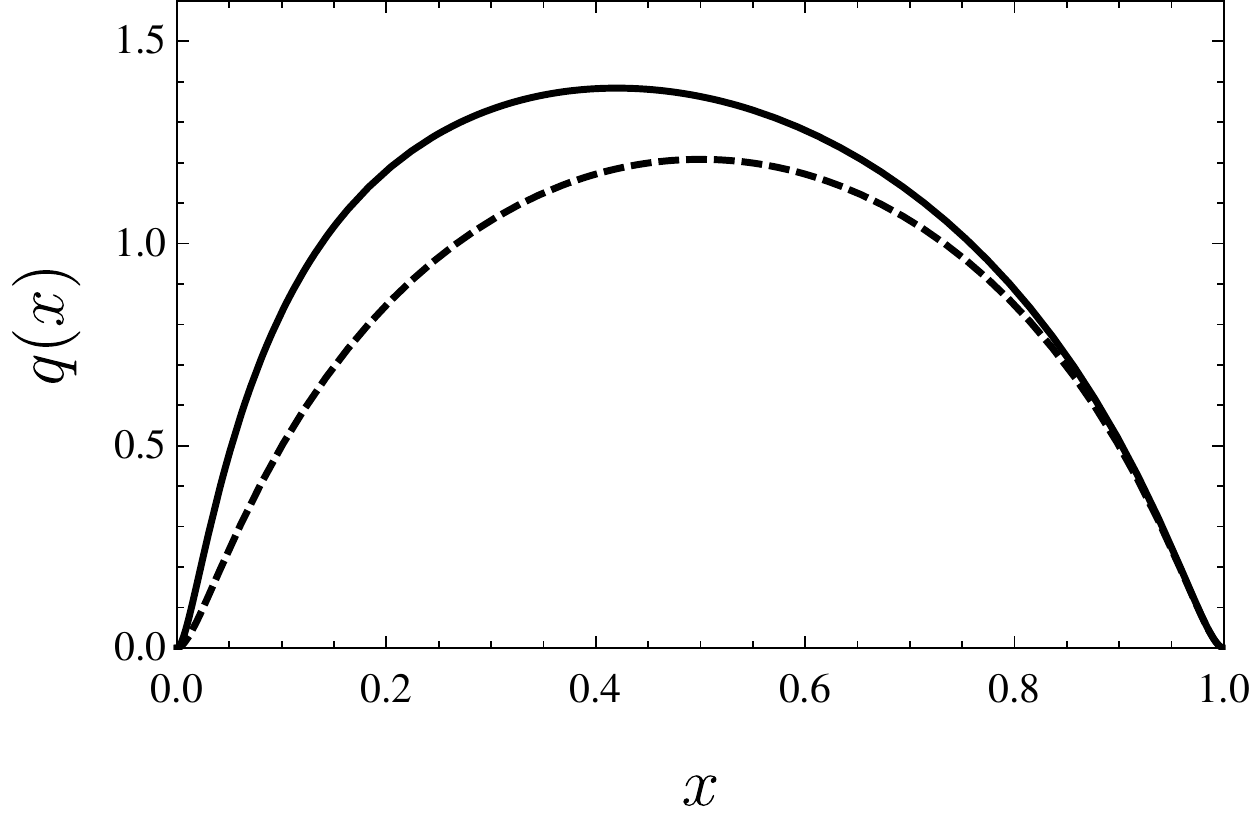}
}
\caption{
Quark distribution function in the pion. 
Using the model Bethe-Salpeter vertex in 
Eq.~\eqref{eq:BSvertex}, 
the normalized quark distribution function, 
$q(x)$,  
is obtained using 
Eq.~\eqref{eq:q}, 
and plotted as a function of 
$x$
(solid curve). 
Additionally shown
(dashed curve)
is the quark distribution function obtained from the two-body light-front wavefunction, 
Eq.~\eqref{eq:LFWF}.
The effect of higher Fock-state contributions in this model reduces the average 
$x$, 
however, 
higher Fock states only account for 
$16 \%$ 
of the probability distribution. 
}\label{f:pdf}
\end{figure}
%
%
%
%%%%%%%%%%%%%%%%%%%%%%%%%%%%%%%%%%%%%%%

Determination of the DDs from the above equation is rather technical, 
however, 
the procedure outlined in 
Refs.~\cite{Tiburzi:2002tq,Tiburzi:2004mh} 
can be adapted to the present model. 
We provide the essential details leading to the extraction of
$F(\b,\a,t)$
and
$G(\b,\a,t)$
from 
Eq.~\eqref{eq:mel}
in Appendix~\ref{s:A}. 
From these DDs, 
we obtain the pion GPD via 
Eq.~\eqref{eq:GPDD}.
To evaluate the GPD, 
we require model parameters. 
In Ref.~\cite{Fanelli:2016aqc}, 
central values for the model parameters are adopted:
$m = 0.220 \, \texttt{GeV}$
and
$m_R = 1.192 \, \texttt{GeV}$. 
These values reproduce the pion decay constant determined within the model, 
and give a good description of the experimentally measured vector form factor of the pion. 
Once the model parameters are fixed, 
the normalization factor 
$\cN^2$
follows from the value of the vector form factor at vanishing momentum transfer,
$F(t=0) = 1$;
or, 
equivalently from normalizing the quark distribution, 
$\int dx \, q(x) = 1$. 
The model's quark distribution is shown in Fig.~\ref{f:pdf}.
This quark distribution is contrasted with that obtained from the two-body light-front wavefunction, 
which will be determined below in 
Sec.~\ref{s:LFWF}. 
One attributes the differences as due to higher Fock components introduced by the covariant Ansatz for the Bethe-Salpeter vertex.

%%%%%%%%%%%%%%%%%%%%%%%%%%%%%%%%%%%%%%%
%
%
%
\begin{figure}
%%%%%%%%%%%%
%%%%%%%%%%%%
\resizebox{\linewidth}{!}{
%%%%%%%%%%%%
%%%%%%%%%%%%
\includegraphics{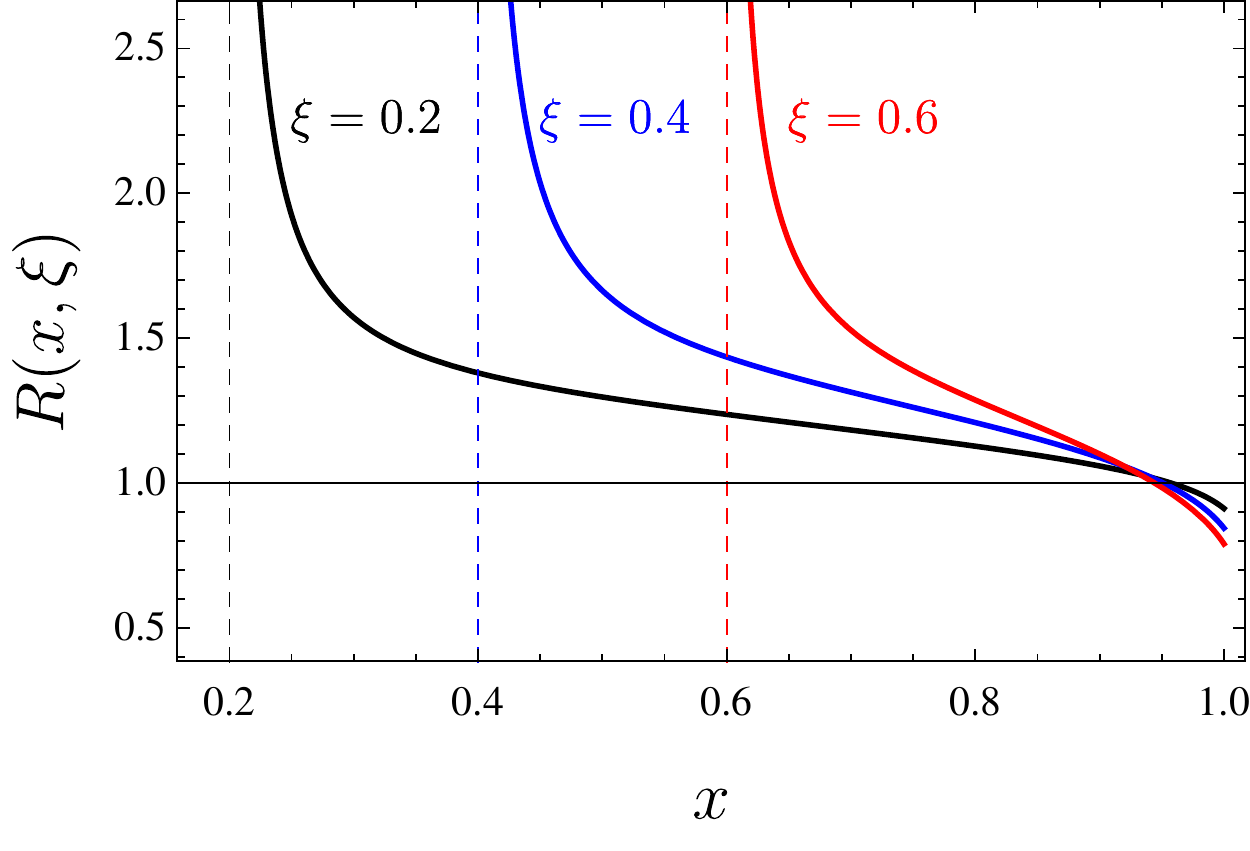}
}
%%%%%%%%%%%%
%%%%%%%%%%%%
\resizebox{\linewidth}{!}{
%%%%%%%%%%%%
%%%%%%%%%%%%
\includegraphics{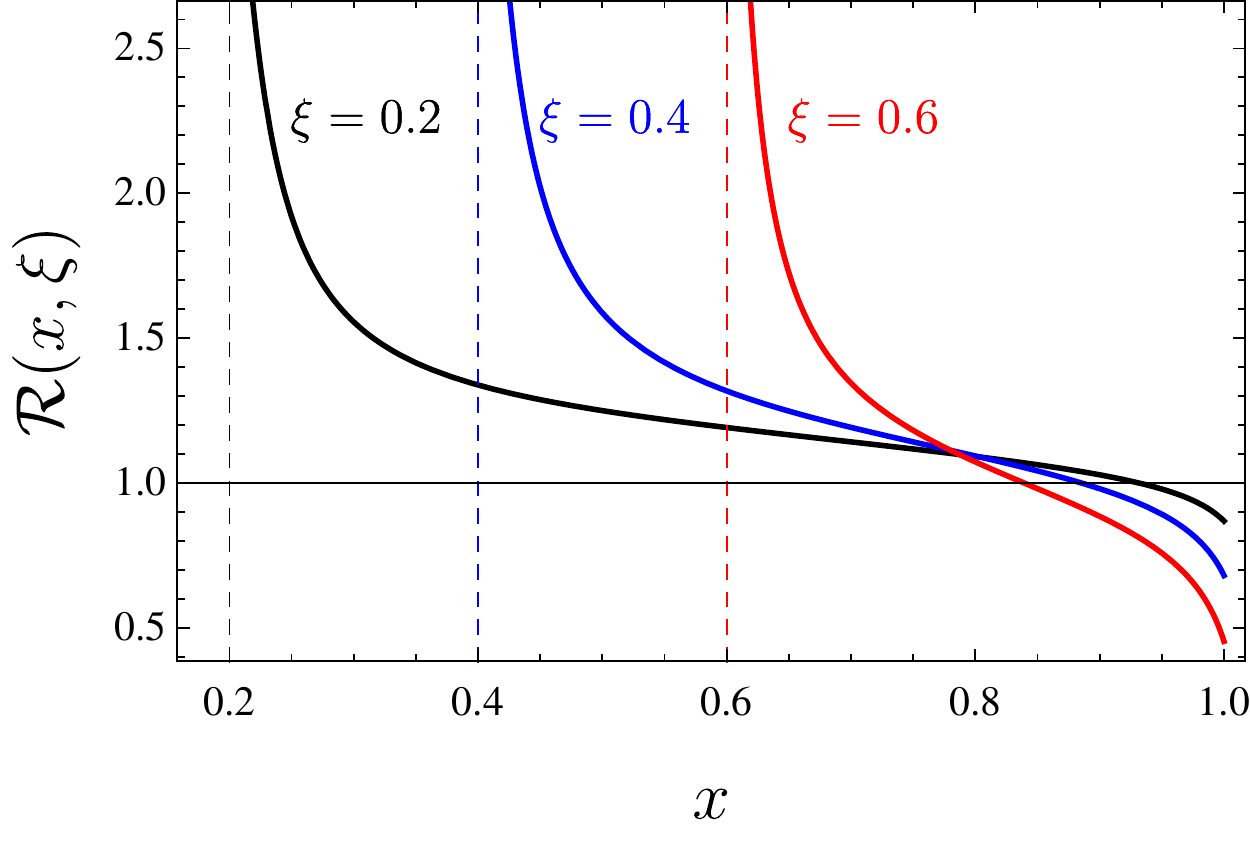}
}
\caption{
Investigation of the positivity bounds required on the GPD within the model.
The top panel shows the ratio 
$R(x,\xi)$ 
in 
Eq.~\eqref{eq:R1}, 
which tests the strongest bound. 
This bound is tested using the listed values of 
$\xi$, 
with the dashed vertical lines showing the corresponding starting points, 
where 
$x = \xi$.
The bottom panel shows the ratio
$\cR(x,\xi)$ appearing in Eq.~\eqref{eq:R2}. 
This ratio tests the weaker positivity bound. 
For 
$x> \xi$, 
both ratios are required to be less than unity. 
}
\label{f:positivity}
\end{figure}
%
%
%
%%%%%%%%%%%%%%%%%%%%%%%%%%%%%%%%%%%%%%%

\begin{figure}
%%%%%%%%%%%%
%%%%%%%%%%%%
\resizebox{\linewidth}{!}{
%%%%%%%%%%%%
%%%%%%%%%%%%
\includegraphics{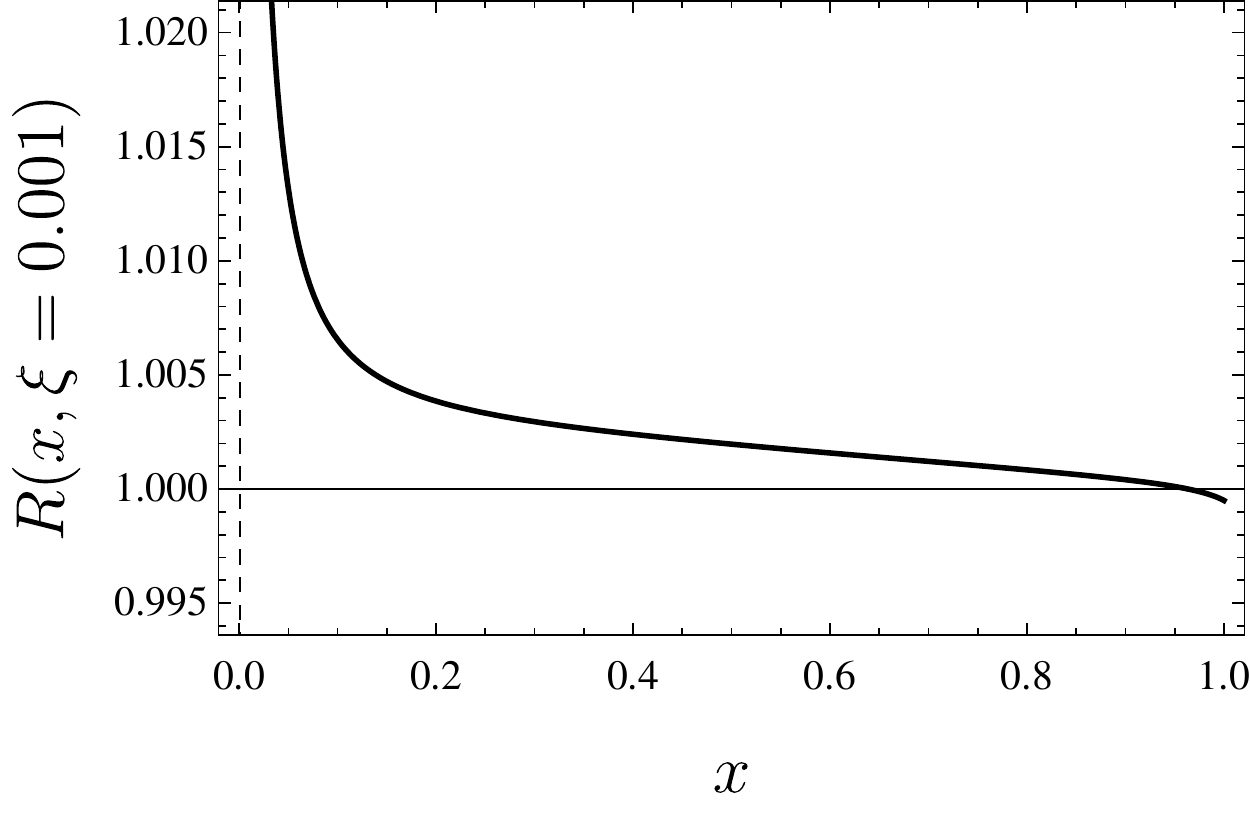}
}
\caption{
Saturation of the (strong) positivity bound in the limit 
$\xi \to 0$. 
The ratio 
$R(x,\xi)$
appearing in 
Eq.~\eqref{eq:R1} 
is plotted as a function of 
$x$
for the skewness
$\xi = 0.001$, 
for which the value 
$x = \xi$
is indicated by the dashed vertical line. 
Notice the considerable reduction in the range plotted compared to that in
Fig.~\ref{f:positivity}. 
Violation of the bound appears linked with the two limits: 
$x \to \xi$
and
$\xi \to 0$. 
}
\label{f:saturate}
\end{figure}

To test the self-consistency of the model, 
we investigate the positivity bound, 
which arises from the positivity of the norm on Fock space. 
The strongest bound arises at vanishing momentum transfer.%
\footnote{
At
$t=0$, 
the consideration of nonzero skewness, 
$\xi \neq 0$,
places one in an unphysical regime,  
which
technically requires analytic continuation. 
In terms of the Lorentz invariant DDs, 
this analytic continuation is trivially carried out by evaluating 
Eq.~\eqref{eq:GPDD}
at 
$t = 0$. 
In the light-front wavefunction overlap representation, 
one first performs the integration over the transverse momentum, 
which produces functions of 
$\bm{\Delta}_\perp^2$, 
and then evaluates at the point
$\bm{\Delta}_\perp^2 = - 4 \xi^2 M^2$.   
We have checked that the two procedures produce identical results. 
Additionally in the \emph{physical} regime, 
with 
$t < 0$ 
and 
$0 < \xi < \left(1 - \frac{4 M^2}{t}\right)^{-1/2}$, 
violation of the positivity bound is qualitatively quite similar to what is shown in Fig.~\ref{f:positivity}. 
}
To this end, 
we form the ratio
\begin{eqnarray}
R(x,\xi) 
\equiv
\frac{|H(x,\xi,0)|}{
\sqrt{  q ( x_f ) q( x_i)}
}
\leq 1
\label{eq:R1}
,\end{eqnarray}
which is bounded by unity from above for
$x>\xi$. 
The momentum fractions carried by the struck quark before and after interacting with the current are given by 
\begin{equation}
x_i = \frac{x+\xi}{1+\xi}, 
\quad
\text{and}
\quad
x_f = \frac{x-\xi}{1-\xi}
.\end{equation}
Investigation of this ratio as a function of 
$x$ 
is shown in Fig.~\ref{f:positivity}
using a few values for the skewness parameter
$\xi$. 
The model of 
Ref.~\cite{Fanelli:2016aqc} 
violates the positivity bound in a window
$\xi < x < x_m(\xi)$, 
encompassing the majority of 
$x$ 
values.
The turn-over point, 
$x_m(\xi) \sim 0.95$,
depends mildly on 
$\xi$
as can be discerned from the figure. 
In the narrow end-point region, 
$x_m(\xi) < x < 1$, 
the bound is satisfied, 
however, 
this region shrinks to zero as the skewness approaches zero. 
Notice that as 
$\xi \to 0$, 
the positivity bound must be saturated according to the limiting behavior of the GPD in 
Eq.~\eqref{eq:q}; 
however, 
the approach is from above rather than below. 
This is exactly the opposite behavior required by the positivity of the norm on Fock space. 
Saturation of the bound is investigated in 
Fig.~\ref{f:saturate}.

%%%%%%%%%%%%%%%%%%%%%%%%%%%%%%%%%%%%%%%
%
%
%
%
%
%
%%%%%%%%%%%%%%%%%%%%%%%%%%%%%%%%%%%%%%%

Due to violation of the positivity bound in 
Eq.~\eqref{eq:R1}, 
we investigate the severity of violation by considering a weaker bound. 
Weaker than the bound involving the geometric mean of quark distributions, 
there is a positivity bound required due to the inequality between geometric and arithmetic means
\begin{equation}
\sqrt{ q (x_f)  q(x_i)}
\leq 
\frac{
q(x_f)
+ 
q(x_i) 
}{2}
.\end{equation}
Hence, 
we consider an additional ratio to test the weaker bound
\begin{equation}
\cR (x,\xi)
=
\frac{|H(x,\xi,0)|}{
\frac{1}{2}
\left[
q(x_f)
+ 
q(x_i) 
\right]
}
\leq 1
\label{eq:R2}
.\end{equation}
This ratio is shown in 
Fig.~\ref{f:positivity} 
for the same values of 
$\xi$. 
Violation of the weaker bound qualitatively follows the same trend as seen in violation of the stronger bound. 
The violation is less severe, 
but that is precisely what is expected of a weaker bound.

Despite the fully covariant nature of the model, 
the positivity bounds required of the GPD are violated. 
This points to the model not being self consistent. 
Without knowledge of the dynamics giving rise to the Bethe-Salpeter vertex function, 
it is not at all obvious how to consistently couple the quark bilocal operator to the dynamics. 
Investigation of the light-front representation for the GPD provides considerable intuition about the shortcomings of the model;
and, 
it is to this representation that we now turn.

%%%%%%%%%%%%%%%%%%%%%%%%%%%%%%%%%%%%%%%
\section{Light-Front Wavefunction Analysis}
\label{s:LFWF}
%%%%%%%%%%%%%%%%%%%%%%%%%%%%%%%%%%%%%%%

%%%%%%%%%%%%%%%%%%%%%%%%%%%%%%%%%%%%%%%
\subsection{Two-Body Analysis}

The two-body wavefunction for the pion is obtained by the overlap of the 
Bethe-Salpeter wavefunction onto on-shell quark and antiquark spinors. 
Subsequent integration over the minus-component of momentum produces the restriction to the plane
$x^+ = 0$, 
and the light-front wavefunction reads, 
see, e.g.~\cite{Liu:1992dg}, 
\begin{eqnarray}
\psi^{(2)}_{\lambda \lambda'}
(x, \bm{k}_\perp)
&=&
\int \frac{dk^-}{2\pi}
\frac{\ol u_\lambda(k^+,\bm{k}_\perp)}{\sqrt{k^+}}
\gamma^+ \Psi(k,P) \gamma^+
\notag \\
&& \phantom{space}
\times 
\frac{v_{\lambda'}(P^+ - k^+,-\bm{k}_\perp)}{\sqrt{P^+ - k^+}}
\label{eq:LFWF2}
,\end{eqnarray}
where we have taken 
$\bm{P}_\perp = \bm{0}$
for simplicity. 
In the wavefunction, 
the variable 
$x$ 
is the plus-momentum fraction of the quark relative to that of the pion, 
$x = k^+ / P^+$. 
Because the transverse boosts generate a kinematic subgroup of the Poincar\'e group, 
the boosted wavefunction, 
$\bm{P}_\perp \neq \bm{0}$, 
can be obtained by the simple replacement
$\bm{k}_\perp \to \bm{k}_\perp - x \bm{P}_\perp$, 
which is the relative transverse momentum.

Using the Bethe-Salpeter wavefunction in Eq.~\eqref{eq:model}, 
the two-body (valence) light-front wavefunction can be obtained. 
For manipulations involving the light-front spinors, 
consult Appendix~\ref{s:B}. 
Treating the factor of 
$\theta[x(1-x)]$
implicitly, 
we arrive at the wavefunction
\begin{eqnarray}
\psi^{(2)}_{\lambda \lambda'} (x,\bm{k}_\perp)
&=&
\cN \,
\frac{
k_{-\lambda} \delta_{\lambda,\lambda'} - \lambda m \,\delta_{\lambda, - \lambda'}
}
{m_R^2 - m^2}
\notag \\
&&
\times
\Big[
\varphi(x,\bm{k}_\perp; m, m, m_R)
\notag \\
&& 
\phantom{s}
-
\varphi(x,\bm{k}_\perp; m_R, m, m_R)
\Big]
,\label{eq:LFWF}
\end{eqnarray}
which has been written with the help of the helicity independent amplitude
\begin{eqnarray}
\varphi(x,\bm{k}_\perp; m_a, m_b, m_c)
&=&
D_W(x,\bm{k}_\perp ; m_a, m_b)
\notag \\
&& \phantom{sp}
\times 
\frac{D_W(x,\bm{k}_\perp ; m_a, m_c)}{x^2 (1-x)}
.\quad
\end{eqnarray}
This amplitude has been expressed in terms of the propagator for the Weinberg equation%
~\cite{Weinberg:1966jm} 
\begin{equation}
D_W(x,\bm{k}_\perp ; m_a, m_b ) 
= 
\frac{1}{
M^2 - \frac{\bm{k}_\perp^2 + x m_a^2 + (1-x) m_b^2}{x(1-x)}
}
,\end{equation}
in the unequal mass case. 
At this stage, 
the first mass argument of the amplitude 
$\varphi$
is redundant, 
however, 
the dependence on three different masses will be utilized below in computing the GPD. 
The light-front helicity structure of the wavefunction in Eq.~\eqref{eq:LFWF} has both opposite helicity, 
$\delta_{\lambda, - \lambda'}$
and same helicity, 
$\delta_{\lambda, \lambda'}$
contributions. 
The latter are accompanied by a unit of quark orbital angular momentum to preserve the spin of the pion. 
This is reflected by the factor of transverse momentum, 
$k_{\lambda} = k_1 + i \lambda k_2$, 
appearing in the numerator of the wavefunction. 
Such contributions are symmetric under the interchange of the mass parameters,
$m \leftrightarrow m_R$. 
The contributions from opposite helicity states are not symmetric under this interchange.

The two-body light-front wavefunction is not symmetric under the interchange of quark and antiquark, 
i.e.~$x \leftrightarrow 1 - x$. 
The wavefunction vanishes linearly at both endpoints, 
$x = 0$
and
$x =1$. 
The vanishing at 
$x = 0$
is particularly interesting because it arises from an exact cancellation between the two terms appearing in 
Eq.~\eqref{eq:LFWF}. 
The quark distribution obtained from the two-body light-front wavefunction takes the simple quantum mechanical form 
\begin{equation}
q^{(2)}(x) 
= 
\sum_{\lambda, \lambda'}
\int \frac{d \bm{k}_\perp}{2 (2\pi)^3}
\left| \psi^{(2)}_{\lambda \lambda'} (x, \bm{k}_\perp) \right|^2
\label{eq:q2}
,\end{equation}
and is plotted in Fig.~\ref{f:pdf}.
The absolute normalization is closely examined in Appendix~\ref{s:B}. 
Qualitatively this two-body quark distribution is similar to the full quark distribution function obtained in the model. 
The full distribution has a smaller value of 
$\langle x \rangle$, 
for example, 
which is consistent with expectations from higher Fock components. 
The difference is not very appreciable: 
$\langle x \rangle = 0.47$ 
in the full model, 
while 
$\langle x \rangle^{(2)} = 0.50$ 
within the two-body sector. 
The 
$x \leftrightarrow 1-x$ 
asymmetry is practically negligible for the two-body quark distribution, 
which is a fortunate circumstance on physical grounds. 
Integrating the two-body quark distribution over 
$x$, 
we find that the two-body Fock state
accounts for 
$84 \%$
of the quark distribution.

From the two-body wavefunction, 
one can obtain the GPD only in a limited kinematic regime. 
In the region 
$x > \xi$, 
we have
\begin{eqnarray}
H^{(2)}(x, \xi, t)
&=& 
%\theta(x - \xi)
\sum_{\lambda, \lambda'}
\int \frac{d \bm{k}_\perp}{2 (2\pi)^3}
\psi^{(2)}_{\lambda \lambda'} \left(x_i, \bm{k}_\perp \right) 
\psi^{(2)*}_{\lambda \lambda'} \left(x_f, \bm{k}'_\perp  \right) 
,\notag \\
\end{eqnarray}
where the boosted transverse momentum is given by
\begin{equation}
\bm{k}'_\perp = \bm{k}_\perp + \left( 1 - \frac{x_i + x_f}{2} \right) \bm{\Delta}_\perp
.\end{equation} 
Due to the convolution of the initial- and final-state wavefunctions, 
the two-body GPD self consistently satisfies the positivity bound. 
The tradeoff is that the two-body approximation violates Lorentz invariance
(for example,  
the sum-rule for the form factor does not hold). 
Additionally when 
$x < \xi$, 
higher Fock components are necessarily required for non-vanishing contributions to the GPD.

%%%%%%%%%%%%%%%%%%%%%%%%%%%%%%%%%%%%%%%
\subsection{Complete Light-Front Analysis}

One similarity between the two-body and higher Fock-state contributions is that they both vanish at 
$x = 0$. 
While this is required of the former, 
the latter are generally non-vanishing at $x=0$. 
Elaboration on this point is made in 
Refs.~\cite{Antonuccio:1997tw,Tiburzi:2002sx}.
The vanishing of this model's quark distribution at 
$x = 0$
leads us to the culprit of positivity violation. 
To demonstrate this, 
we turn to the light-front representation of the model's GPD.

Complementary to the representation in terms of DDs, 
the GPD can be evaluated directly from the bilocal matrix element in 
Eq.~\eqref{eq:bilocal}
computed within the model. 
This alternative expression for the GPD is 
\begin{eqnarray}
H(x,\xi,t)
&=&
\frac{1}{2 \ol P {}^+}
\int_k
\delta \left( x + \xi - \frac{k^+ }{\ol P {}^+} \right)
\,
\Tr \Big[ \Psi(k,P)
\notag \\
&& \phantom{spa} \times
 S^{-1}(k-P) \ol \Psi(k', P') \gamma^{+} \Big]
\label{eq:LCGPD}
.\end{eqnarray}
Results for the quark distribution and GPD obtained from 
Eq.~\eqref{eq:LCGPD} 
agree numerically with those determined from DDs. 
Our investigation of the violation of the positivity bound leads us to consider the GPD at the crossover point, 
$x = \xi$. 
We show in the form of an integral that the GPD is nonvanishing at the crossover, 
$H(\xi, \xi, t) \neq 0$, 
using light-front integration.%
\footnote{
Given that the quark distribution vanishes at the endpoint, 
$x=0$,
it is already numerically evident from 
Figs.~\ref{f:positivity} and \ref{f:saturate} 
that the GPD does not vanish at the crossover.} 
This result can also be established analytically from the DDs, 
however, 
the light-front integration offers a more intuitive 
picture, 
not to mention the ability to display using more compact expressions.

To obtain the GPD, 
we integrate the expression in Eq.~\eqref{eq:LCGPD} over 
$k^-$. 
The integral is performed by the residue theorem, 
and different residues are required depending on the value of 
$k^+$.  
A particularly illustrative discussion of the contour integration for light-front and instant-form dynamics can be found in 
Ref.~\cite{Sawicki:1991sr}. 
We restrict our attention to the region
$x > \xi$, 
for which residues at the spectator poles are required. 
There are two such poles: 
one for the spectator quark, 
and one for the spectator regulator particle having mass 
$m_R$. 
Due to the Ansatz in 
Eq.~\eqref{eq:BSvertex}, 
the spectator propagator for the regulator is squared in the bilocal current matrix element. 
For ease in evaluating and displaying the results, 
we use the standard trick to rewrite the square of a propagator in terms of a derivative of the propagator, 
namely
\begin{equation}
\frac{1}{\left[(P-k)^2 - m_R^2\right]^2} 
= 
\frac{\partial}{\partial m_A^2}
\frac{1}{(P-k)^2 - m_A^2} 
\Bigg|_{m_A = m_R}
.\end{equation}
A different mass, 
$m_A$, 
is required for this intermediate step because 
$m_R$
appears elsewhere in the expression for the GPD. 
Carrying out the light-front integration, 
we arrive at the result
\begin{eqnarray}
\theta(x - \xi)
H(x,\xi,t)
= 
\cN^2
\int \frac{d \bm{k}_\perp}{2 (2 \pi)^3}
\frac{\partial}{\partial m_A^2}
\frac{\cH_m - \cH_A}
{m^2 - m_A^2}
\Bigg|_{m_A = m_R}
\notag \\
\label{eq:LFGPD}
\end{eqnarray}
where 
$\cH_m$
and
$\cH_A$
denote contributions to the integrand that arise from taking the residue at the spectator quark pole
and spectator regulator pole, 
respectively. 
The former is given by
\begin{eqnarray}
\cH_m
&=&
2 \left( \bm{k}_\perp \cdot \bm{k}'_\perp + m^2 \right)
\notag \\
&& 
\times
\varphi(x_i, \bm{k}_\perp; m, m_R, m)
\,
\varphi(x_f, \bm{k}'_\perp; m, m_R, m)
, \quad\end{eqnarray}
while the latter takes the form
\begin{eqnarray}
\cH_A
&=&
2 \left( \bm{k}_\perp \cdot \bm{k}'_\perp + \frac{1-x^2}{1-\xi^2} m^2 + x_i \, x_f \, m_A^2 \right)
\notag \\
&& 
\times
\varphi(x_i, \bm{k}_\perp; m_A, m, m_R)
\,
\varphi(x_f, \bm{k}'_\perp; m_A, m, m_R)
.\notag \\
\end{eqnarray}

Using Eq.~\eqref{eq:LFGPD}, 
we derive the crossover behavior by taking the limit
$x \to \xi$
from above. 
To this end, 
it is useful to note the finite limit
\begin{equation}
\lim_{x \to 0}
\varphi(x, \bm{k}_\perp; m_a, m_b, m_c)
=
\frac{1}{(\bm{k}_\perp^2 + m_b^2) (\bm{k}_\perp^2 + m_c^2)}
\label{eq:lemma}
,\end{equation}
which is independent of the mass 
$m_a$. 
As a consequence, 
we have the non-vanishing result
%%%%%%%%%%%%%%%%%%%%%%%%%%%%%%%%%%%%%%%
\begin{widetext}
\begin{eqnarray}
H(\xi, \xi, t)
&=&
\cN^2
\int \frac{d \bm{k}_\perp}{(2 \pi)^3}
\frac{\bm{k}_\perp \cdot \bm{k}^{(0)}_\perp + m^2 }
{\left[\left(\bm{k}^{(0)}_\perp\right)^2 + m^2\right]
\left[ \left(\bm{k}^{(0)}_\perp\right)^2 + m_R^2\right]}
%\notag \\
%&&
%\times 
\,
\frac{\partial}{\partial m_A^2}
\frac{\varphi(x_0, \bm{k}_\perp; m, m_R, m)
 - \varphi(x_0, \bm{k}_\perp; m_A, m, m_R)
}{m^2 - m_A^2}
\Big|_{m_A = m_R},
\notag \\
\label{eq:cross}
\end{eqnarray}
\end{widetext}
%%%%%%%%%%%%%%%%%%%%%%%%%%%%%%%%%%%%%%%
where we have defined the crossover values
\begin{eqnarray}
x_0 
&=&
x_i(x = \xi )
=
\frac{2 \xi}{1 + \xi}
,\end{eqnarray}
for the initial quark's momentum fraction, 
and
\begin{eqnarray}
\bm{k}_\perp^{(0)}
&=&
\bm{k}_\perp + \frac{1}{1 + \xi} \bm{\Delta_\perp}
,\end{eqnarray}
for the relative transverse momentum of the final-state quark. 
Finally, 
taking the limit 
$\xi \to 0$
results in 
\begin{equation}
\lim_{\xi \to 0} 
H(\xi, \xi, t)
= 
0
,\end{equation}
which can be easily demonstrated upon noting that
$x_0 \to 0$
in this limit, 
then subsequently applying 
Eq.~\eqref{eq:lemma} to the expression for the GPD at the crossover, 
Eq.~\eqref{eq:cross}.% 
\footnote{
One can generalize these findings in the following way. 
Adopting a more general 
(but asymmetric)
Ansatz for the Bethe-Salpeter amplitude having the form
\begin{equation}
\Gamma(k,P)
=
\cN \, 
[k^2 - m_R^2 + i \e]^{-1}[(P-k)^2 - m_R^2 + i \e]^{-\nu / 2}
,\notag \end{equation}
for 
$\nu = 1$, $2$, $\cdots$, 
it is straightforward to show using light-front integration that the corresponding crossover value of the GPD, 
$H(\xi, \xi, t)$, 
is non-vanishing, 
but vanishes in the limit of vanishing skewness, 
$\xi \to 0$. 
Thus models of this type will violate the positivity bound in an analogous way to the 
$\nu = 2$ 
model detailed above.
The model with 
$\nu =0$, 
by contrast, 
satisfies the positivity bound, 
and was considered previously in 
Refs.~\cite{Tiburzi:2002tq,Tiburzi:2004mh}.
This particular case is best viewed as smearing the bilocal current rather than the bound-state vertex, 
see Ref.~\cite{Bakker:2000pk}. 
}

%%%%%%%%%%%%%%%%%%%%%%%%%%%%%%%%%%%%%%%
%
%
%
\begin{figure}
%%%%%%%%%%%%
%%%%%%%%%%%%
\resizebox{\linewidth}{!}{
%%%%%%%%%%%%
%%%%%%%%%%%%
\includegraphics{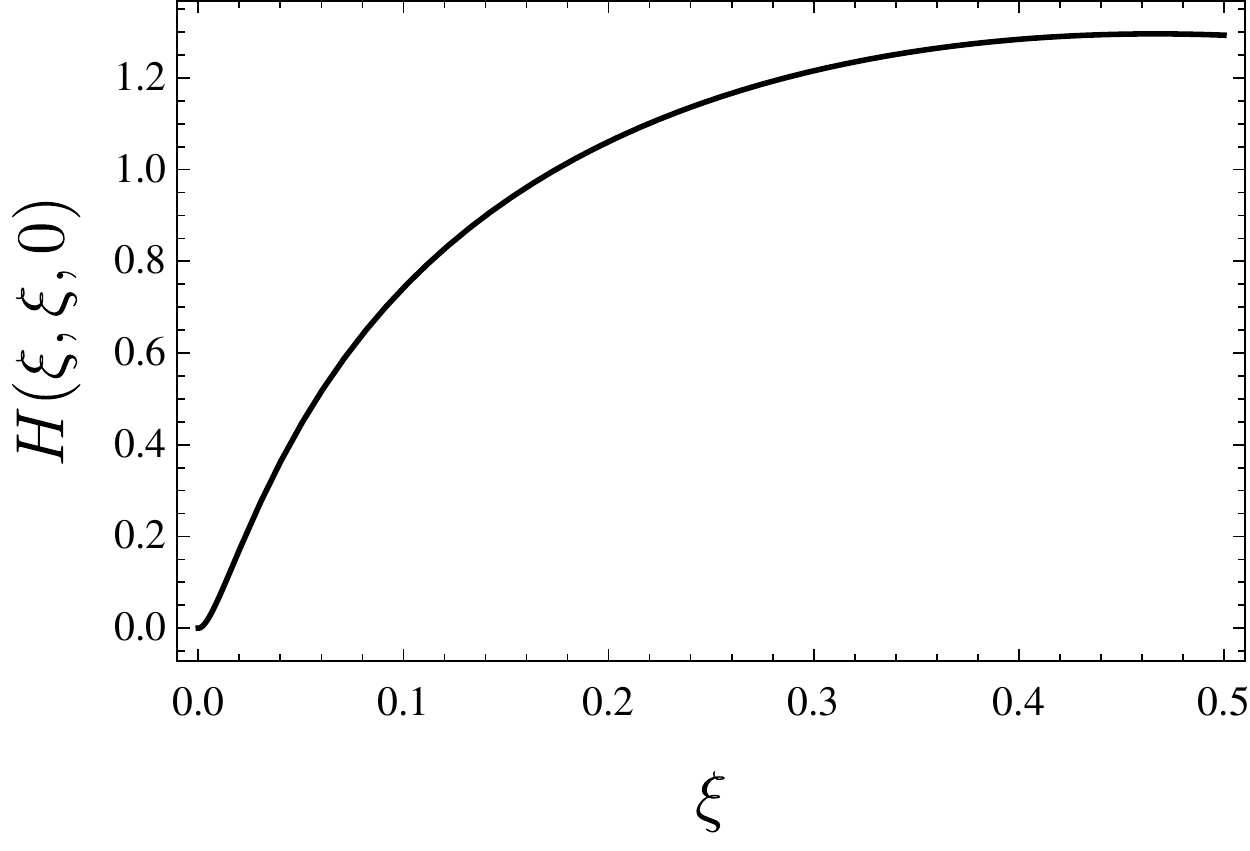}
}
\caption{
Value of the GPD at the crossover point, 
$x = \xi$, 
plotted as a function of 
$\xi$. 
For simplicity, 
we determine the value using 
$t = 0$. 
The vanishing 
$\xi \to 0$
limit of the crossover value, 
implies vanishing of the quark distribution function at the endpoint, 
$x = 0$.  
The non-vanishing value of 
$H(\xi, \xi, t)$
for 
$\xi \neq 0$, 
then leads to (infinite) violation of the positivity bound in Eq.~\eqref{eq:R1}.
Close inspection shows that 
$H(\xi, \xi, t) \propto \xi^2$, 
for
$\xi \ll 1$. 
}
\label{f:cross}
\end{figure}
%
%
%
%%%%%%%%%%%%%%%%%%%%%%%%%%%%%%%%%%%%%%%

A plot of the 
$\xi$-dependence of the GPD at the crossover is shown in Fig.~\ref{f:cross}, 
and confirms its non-vanishing value, 
as well as the vanishing forward limit. 
These two features are directly at odds with the positivity of the norm on the light-front Fock space. 
The vanishing of the crossover's forward limit confirms that the model's quark distribution function vanishes at 
$x = 0$, 
as we have numerically shown above.
In terms of the general light-front wavefunction representation, 
we have schematically
\begin{equation}
q(x = 0) 
= 
H(0,0,0)
=
\sum_n
\Big| \psi^{(n)} (x = 0, \cdots) \Big|^2
=0
.\end{equation}
We write the $n$-body Fock component of the bound state as 
$\psi^{(n)}$, 
and only indicate the active quark's momentum fraction 
$x$, 
which is at the endpoint, 
$x = 0$. 
Due to positivity of the norm, 
the vanishing quark distribution at the endpoint implies that 
\begin{equation}
\psi^{(n)} (x = 0, \cdots) = 0, 
\quad 
\forall n
\label{eq:end}
.\end{equation}
Approaching the crossover from above, 
the GPD at the crossover has an expansion in terms of diagonal Fock-component overlaps,
which schematically takes the form
\begin{eqnarray}
H(\xi,\xi,t)
= 
\sum_n 
\psi^{(n)}( x_i, \cdots)
\psi^{* (n)} (x_f = 0, \cdots ) 
.\end{eqnarray}
Given the endpoint behavior of the wavefunctions deduced from the norm, 
Eq.~\eqref{eq:end}, 
we should very likely have 
$H(\xi, \xi, t) = 0$
for self consistency, 
which is in contradiction with 
Eq.~\eqref{eq:cross}.%
\footnote{
Mathematically speaking, 
it is possible that the crossover value,  
as the sum of infinitely many vanishing contributions, 
need not itself vanish. 
If this scenario is realized, 
however, 
the positivity bound in 
Eq.~\eqref{eq:R1}
will nevertheless be violated. 
} 
This contradiction is precisely the source of the observed (infinite) violation of the positivity bound in 
Eq.~\eqref{eq:R1} 
at 
$x = \xi$. 
Away from the crossover, 
violation then persists until the relatively large value of 
$x = x_m(\xi)$, 
as discussed above.

%%%%%%%%%%%%%%%%%%%%%%%%%%%%%%%%%%%%%%%
\section{Summary}
\label{s:summy}
%%%%%%%%%%%%%%%%%%%%%%%%%%%%%%%%%%%%%%%

We investigate GPDs in the quark model for the pion proposed in 
Ref.~\cite{Fanelli:2016aqc}. 
A feature of this model is the covariant smearing of the Bethe-Salpeter vertex, 
which was put forward in order to provide realistic bound-state structure beyond the point-like approximation. 
The covariant nature of the model, 
moreover, 
guarantees that all constraints associated with Lorentz invariance of the underlying matrix element are automatically satisfied. 
This fact enables us in 
Sec.~\ref{s:DD} 
to derive the DDs for the model, 
see 
Eq.~\eqref{eq:FandG}. 
The GPD obtained from the DDs is then scrutinized using the ratios in 
Eqs.~\eqref{eq:R1} and \eqref{eq:R2}. 
Positivity of the norm on the light-front Fock space requires both of these ratios to be less than unity, 
with the former bound stronger than the latter. 
Both bounds, 
however, 
are violated for nearly all values of 
$x > \xi$. 
This is shown in 
Fig.~\ref{f:positivity}. 
Such violation, 
moreover, 
is tied to a mismatch between the endpoint behavior of the quark distribution and the crossover value of the GPD, 
see Eq.~\eqref{eq:cross}. 
As the former vanishes and the latter is non-vanishing, 
the mismatch leads to an infinite violation of the positivity bound.

The
GPDs encompass physics of both quark distributions and form factors. 
We have shown that fully covariant models, 
which can be tuned to reproduce the experimentally measured quark distributions and form factors, 
need not give a good description of GPDs. 
In fact, 
the behavior of the model's GPD at the crossover is inconsistent with its quark distribution. 
While the model is defined at a low scale, 
QCD radiation will drive the model's quark distribution at the endpoint away from zero as the renormalization scale is raised.  
The positivity bounds, 
however,  
are stable with respect to evolution. 
Thus, 
the defect in the model at a low scale will persist at higher scales. 
The value of the GPD at the crossover, 
moreover, 
directly enters the imaginary part of the deeply virtual Compton scattering amplitude. 
This imaginary part then appears in the cross section through interference terms with Bethe-Heitler processes, 
which enable GPDs to be accessed experimentally. 
The model therefore fails to describe GPDs at the most phenomenologically relevant point. 
The mismatch in endpoint and crossover behavior requires proper treatment of higher Fock states in order to resolve. 
Such treatment requires one to go beyond the impulse approximation. 
While theoretically challenging, 
consistent modeling of GPDs presents an essential test in the phenomenological description of hadron structure.
Passing this test will help enable great insight into the nature of bound states in QCD.

%%%%%%%%%%%%%%%%%%%%%%%%%%%%%%%%%%%%%%%
\begin{acknowledgments}
%%%%%%%%%%%%%%%%%%%%%%%%%%%%%%%%%%%%%%%

This work was supported in part by the U.S.~National Science Foundation, under Grant No.~PHY15-15738.
BCT would like to thank G.~A.~Miller for numerous insights in studying GPDs.

%%%%%%%%%%%%%%%%%%%%%%%%%%%%%%%%%%%%%%%
\end{acknowledgments}
%%%%%%%%%%%%%%%%%%%%%%%%%%%%%%%%%%%%%%%

%%%%%%%%%%%%%%%%%%%%%%%%%%%%%%%%%%%%%%%
\appendix
%%%%%%%%%%%%%%%%%%%%%%%%%%%%%%%%%%%%%%%

%%%%%%%%%%%%%%%%%%%%%%%%%%%%%%%%%%%%%%%
\section{Derivation of the Double Distributions}
\label{s:A}
%%%%%%%%%%%%%%%%%%%%%%%%%%%%%%%%%%%%%%%

In this Appendix, 
we calculate the pion matrix element of the twist-two operators, 
$\cM^{\mu \mu_1 \cdots \mu_n}$, 
in the model specified by the Bethe-Salpeter vertex in Eq.~\eqref{eq:BSvertex}. 
Thereby, 
we extract the DDs, 
$F(\b,\a,t)$
and
$G(\b,\a,t)$
appearing in Eq.~\eqref{eq:FG}. 
These are utilized in the main text to put the model under scrutiny.

The key observation to obtain the DDs from 
Eq.~\eqref{eq:mel}
is that the required binomial coefficients can readily be obtained after reducing momentum factors appearing in the numerator. 
Such factors arise due to the trace over spinor indices, 
which appear in the numerator as
\begin{equation}
N^\mu 
= 
\Tr 
\left[ 
(\rlap \slash k + m)
\gamma_5 
(\rlap \slash k - \Pslash + m) 
\gamma_5 
( \rlap \slash k ' + m )
\gamma^\mu 
\right]
\label{eq:numerator}
.\end{equation}
The trace can be uniquely expressed in terms of factors appearing in the denominator.  
In terms of the momentum dependence, 
there are three such factors due to the inverse propagators, 
which we write as
\begin{eqnarray}
\cA_a 
&=& 
(k-P)^2 - m_a^2 + i \e,
\notag \\
\cB_b 
&=&
(k+\D)^2 
- m_b^2 + i \e,
\notag \\
\cC_c
&=&
k^2 - m_c^2 + i \e
\label{eq:props}
.\end{eqnarray}
These definitions allow us to refer to the constituent quark mass and regulator mass as 
$\cA$ and $\cA_R$, 
respectively. 
It will sometimes be convenient to write 
$\cA_m$ 
for the former. 
Due to its dependence on only the constituent quark mass, 
the trace can be written as
\begin{eqnarray}
N^\mu
&=&
2 \ol P {}^\mu 
\left( t - \cB - \cC \right)
+ 
\D^\mu \left( \cB - \cC \right)
\notag \\
&& \phantom{sp}
+ 
4 \left( k + \D / 2 \right)^\mu \left( M^2 - t/2 - \cA \right)
\label{eq:num}
.\end{eqnarray}
Consequently there are four types of contributions to consider: 
those without a momentum reduction, 
and the three momentum reductions corresponding to cancelling
$\cA$, 
$\cB$, 
and
$\cC$.

In terms of the quantities defined in Eq.~\eqref{eq:props}, 
the momentum denominators of 
Eq.~\eqref{eq:mel} take the form 
\begin{eqnarray}
\frac{1}{\cA \, \cA_R^2 \, \cB \, \cB_R \, \cC \, \cC_R}
=
\frac{\partial}{\partial m_A^2}
\frac{1}{\cA \, \cA_A \, \cB \, \cB_R \, \cC \, \cC_R}
\Big|_{m_A = m_R}
.\quad\,\,\end{eqnarray}
Notice that there are three factors from constituent quark propagators in the above expression, 
$\cA \, \cB \, \cC$, 
which reflect those appearing in the triangle diagram, 
Fig.~\ref{f:triangle}.
The factors from propagators of the regulator particle with mass 
$m_R$
appropriately reflect those resulting from the initial-state Bethe-Salpeter vertex, 
$\cA_R \, \cC_R$, 
as well as the final state, 
$\cA_R \, \cB_R$. 
This, of course, 
leads to the squaring of 
$\cA_R$, 
which is adeptly handled by the differentiation and subsequent evaluation that is shown above.  
Products of two propagators with identical momenta, 
such as 
$\cB \, \cB_R$, 
can be written as differences between single propagators, 
for example
\begin{equation}
\frac{1}{\cB \, \cB_R}
=
\frac{1}{m^2 - m_R^2}
\left( 
\frac{1}{\cB}
- 
\frac{1}{\cB_R} 
\right)
.\end{equation}
It will be convenient to define the multiplicative factor
\begin{equation}
\mu_j = \frac{1}{m^2 - m_j^2}
,\end{equation}
with 
$j = A$ 
and 
$j = R$
as the two cases required below. 
In light of such relations, 
we can turn all products of multiple propagators into various instances of an elementary product of three propagators. 
As an example of this, 
we write the following product as
\begin{eqnarray}
\frac{1}{\cA \, \cA_A \, \cB \, \cB_R \, \cC \, \cC_R}
&=&
\frac{1}{\cA_a \, \cB_b \, \cC_c}
\mu_A \, \mu_R^2 \left( \d_{a,m} - \d_{a, A} \right)
\notag \\
&& \phantom{sp} \times
\left( \d_{b,m} - \d_{b,R} \right)
\left( \d_{c,m} - \d_{c,R} \right)
\label{eq:nonreduced}
,\qquad\end{eqnarray}
where we treat 
$a$, 
$b$, 
and 
$c$ 
as indices that keep track of the required masses in the various contributions.

At this point, 
it is useful to completely detail one particular contribution to the model's DDs, 
with the understanding that the general procedure is quite similar for all contributions. 
From the trace appearing in the numerator, 
Eq.~\eqref{eq:num}, 
we focus on the term proportional to 
$M^2$, 
which does not require a momentum reduction. 
As a result, 
this term depends on (a derivative of) the propagators appearing in Eq.~\eqref{eq:nonreduced}. 
Up to multiplicative constants, 
all contributions from this terms are of the form
\begin{eqnarray}
\d \cM^{\mu \mu_1 \cdots \mu_n}_{abc}
&=&
%\int\frac{d^4k}{(2\pi)^4} 
\int_k
\frac{i}{\cA_a \, \cB_b \, \cC_c} 
4 \, (k+\D/2)^{\big\{\mu}
\notag \\
&& \times 
(k+\D/2)^{\mu_1} \cdots (k+\D/2)^{\mu_n\big\}}
\label{eq:demo}
.\end{eqnarray}
Evaluation of the momentum integral can be carried out using Feynman parameterization. 
With malice aforethought, 
the Feynman parameters are chosen so that the 
$\a$, 
$\b$
variables of the DDs can be identified.  
To this end, 
notice that we can write
\begin{eqnarray}
\frac{1}{\cA_a \, \cB_b \, \cC_c} 
=
\int_0^1 d\b \int_{-1+\b}^{+1-\b} d\a \, 
\frac{1}{\cD_{abc}^3}
,\end{eqnarray}
where the denominators have been combined according to the recipe
\begin{eqnarray}
\cD_{abc} 
=
\b \cA_a + \frac{1}{2} \left( 1 + \a - \b \right) \cB_b + \frac{1}{2} \left( 1 - \a - \b \right) \cC_c
.\notag \\
\end{eqnarray}
In terms of the shifted momentum variable $\ell^\mu$, 
given by
$\ell^\mu = \left[ k - \beta \ol P + (1+ \a ) \D / 2 \right]^\mu$, 
the combined denominator takes the form
\begin{eqnarray}
\cD_{abc}
&=&
\ell^2
- 
D_{abc} 
+ 
i \e
\end{eqnarray}
where
\begin{eqnarray}
D_{abc}
&=&
\b \, m_a^2 + \frac{1}{2} ( 1 + \a - \b) \, m_b^2 + \frac{1}{2} ( 1 - \a - \b) \, m_c^2
\notag \\
&& 
- 
\b( 1-\b) M^2
- 
[(1-\b)^2 - \a^2 ] \, \frac{t}{4}
.
\end{eqnarray}
As a result of the momentum shift, 
this particular contribution to the twist-two matrix element is given by 
\begin{eqnarray}
\d \cM_{abc}^{\mu \mu_1 \cdots \mu_n}
&=&
\int_0^1 d\b \int_{-1+\b}^{+1-\b} d\a  
\int_\ell \frac{4 i}{[\ell^2 - D_{abc} + i \e]^3}
\notag \\
&&
\times
\left( \b \ol P - \a \D / 2 \right)^{\big\{ \mu}
\left( \b \ol P - \a \D / 2 \right)^{\mu_1}
\notag \\
&&
\phantom{space}
\times \cdots
\left( \b \ol P - \a \D / 2 \right)^{\mu_n \big\}}
,\end{eqnarray}
where we have appealed to Lorentz invariance and the traceless property of the tensor. 
The four-momentum integration can be performed, and the binomial expansion carried out. 
This procedure allows for the direct identification of contributions to the 
$F$- 
and 
$G$-type 
DDs. 
For the contribution in 
Eq.~\eqref{eq:demo}, 
we thus find
\begin{eqnarray}
\d F(\b, \a, t)
&=&
\frac{1}{(4\pi)^2}
\frac{\b}{D_{abc}},
\quad 
\d G(\b, \a, t)
=
\frac{1}{(4\pi)^2}
\frac{\a}{D_{abc}}
.
\notag \\
\label{eq:dFdG}
\end{eqnarray}

Having spelled out the necessary steps to obtain the DDs, 
we carry out this procedure on each of the four contributions required by the momentum dependence of the numerator appearing in 
Eq.~\eqref{eq:num}. 
Reducing the numerator appropriately, 
forming propagator differences, 
and then carrying out the Feynman parameterization and momentum integration 
produces the DDs. 
Writing the $F$- and $G$-type DDs in the form of a column vector, 
we find
\begin{widetext}
\begin{eqnarray}
\renewcommand*{\arraystretch}{1.25}
\begin{pmatrix}
F(\b,\a,t)\\
G(\b,\a,t)
\end{pmatrix}
&=&
\frac{\cN^2 \mu_R}{(4 \pi)^2}
\frac{\partial}{\partial m_A^2}
\frac{1}{D_{abc}}
\Bigg[
\mu_R 
\left( \d_{b,m} - \d_{b,R} \right)
\left( \d_{c,m} - \d_{c,R} \right)
\Bigg(
\mu_A
\left( \d_{a,m} - \d_{a, A} \right)
\renewcommand*{\arraystretch}{1.25}
\begin{pmatrix}
 \b (M^2 - \frac{t}{2}) + \frac{t}{2}\\
 \a (M^2 - \frac{t}{2}) \phantom{+its}
\end{pmatrix}
-
\d_{a,A}
\renewcommand*{\arraystretch}{1.25}
\begin{pmatrix}
\b  \\
\a
\end{pmatrix}
\Bigg)
\notag \\
&&
-
\frac{\mu_A}{2}
\left( \d_{a,m} - \d_{a, A} \right)
\Bigg(
\delta_{b,R} 
\left( \d_{c,m} - \d_{c,R} \right)
\renewcommand*{\arraystretch}{1.25}
\begin{pmatrix}
1\\
1
\end{pmatrix}
+
\left( \d_{b,m} - \d_{b,R} \right)
\delta_{c,R}
\renewcommand*{\arraystretch}{1.25}
\begin{pmatrix}
\phantom{-}1 
\\
-1
\end{pmatrix}
\Bigg)
\Bigg]_{m_A = m_R}
\label{eq:FandG}
.\end{eqnarray}
\end{widetext}
In the above expression, 
the first grouping of terms in the first line arises from the non-reduced contribution, 
while the second grouping in the first line arises from the $\cA$-reduced contribution. 
The second line is the sum of $\cB$-reduced and $\cC$-reduced contributions. 
Notice that under the sign reversal of $\a$, 
we have
$D_{abc}(\b,-\a,t) = D_{acb}(\b,\a,t)$. 
Because the contributions to the DDs on the first line are symmetric under the interchange 
$b \leftrightarrow c$, 
these terms produce even and odd contributions to the $F$- and $G$-type DDs, respectively. 
The oddness of contributions to 
$G(\b,\a,t)$ 
is due to the overall factor of 
$\a$
in the first line.  
In the second line, 
the contributions to $F(\b,\a,t)$ are even under $b \leftrightarrow c$, 
while those for $G(\b, \a, t)$ are odd under this interchange. 
Thus the $\a$-symmetry of the model's DDs is consistent with the required time-reversal invariance properties.

%%%%%%%%%%%%%%%%%%%%%%%%%%%%%%%%%%%%%%%
\section{Conventions and Light-Front Spinors}
\label{s:B}
%%%%%%%%%%%%%%%%%%%%%%%%%%%%%%%%%%%%%%%

In this Appendix, 
we make explicit the conventions employed above.  
This includes those for the light-front spinors and the normalization leading to the two-body wavefunction, 
which is detailed using a simple model with a point-like Bethe-Salpeter vertex.

%%%%%%%%%%%%%%%%%%%%%%%%%%%%%%%%%%%%%%%
\subsection{Spinors}
%%%%%%%%%%%%%%%%%%%%%%%%%%%%%%%%%%%%%%%

For any Lorentz four-vector, 
$a^\mu$, 
we define its light-front components as 
$a^\pm = \frac{1}{\sqrt{2}} ( a^0 \pm a^3)$, 
along with 
$\bm{a}_\perp = (a^1, a^2)$. 
Similarly, 
for Dirac matrices
$\gamma^\mu$, 
we have 
$\gamma^\pm  = \frac{1}{\sqrt{2}} ( \gamma^0 \pm \gamma^3)$, 
and 
$\bm{\gamma}_\perp = ( \gamma^1, \gamma^2)$. 
The former can be used to construct the Hermitian projection matrices 
$\Lambda_\pm = \frac{1}{2} \gamma^\mp \gamma^\pm$, 
which satisfy the usual properties:
$\Lambda_+ + \Lambda_- = 1$, 
along with
$(\Lambda_\pm)^2 = \Lambda_\pm$, 
and
$\Lambda_\pm \Lambda_\mp = 0$. 
It is also convenient to employ the Hermitian matrices 
$\b = \gamma^0$, 
and 
$\bm{\alpha}_\perp = \gamma^0 \bm{\gamma}_\perp$. 
Manipulations of light-front spinors are made more economical by the use of the identities:
$\beta \Lambda_\pm = \Lambda_\mp \beta$
and
$\bm{\alpha}_\perp \Lambda_\pm = \Lambda_\mp \bm{\alpha}_\perp$.

In the main text, 
we utilize light-front spinors for quarks and antiquarks. 
These are solutions to the Dirac equation 
and its conjugate
\begin{eqnarray}
\left( \rlap \slash k_\text{on} - m \right) u_\lambda (k^+, \bm{k}_\perp) 
&=& 
0
,\notag \\
\left( \rlap \slash k_\text{on} + m \right) v_\lambda(k^+, \bm{k}_\perp) 
&=&
0
,\end{eqnarray}
where the light-front energy is on shell, 
$k^-_\text{on} = \frac{\bm{k}_\perp^2 + m^2}{2 k^+}$. 
Such solutions can be written in terms of the Dirac basis spinors 
\begin{equation}
X_+ = \frac{1}{\sqrt{2}} \begin{pmatrix} 1 \\ 0 \\ 1 \\ 0 \end{pmatrix}, 
\quad 
X_- = \frac{1}{\sqrt{2}} \begin{pmatrix} \phantom{-} 0 \\ \phantom{-}1 \\  \phantom{-}0 \\ -1 \end{pmatrix}
,\end{equation}
which are eigenstates of the projector onto the so-called good spinor components, 
$\Lambda_+ X_\lambda = X_\lambda$. 
Explicit solutions for the light-front spinors are
\begin{eqnarray}
u_\lambda (k^+, \bm{k}_\perp) 
&=&
\frac{1}{(\sqrt{2} k^+)^{\frac{1}{2}}}
\left[
\sqrt{2} k^+
+ 
\beta m
+ 
\bm{\alpha}_\perp \cdot \bm{k}_\perp
\right]
X_\lambda,
\notag \\
v_\lambda (k^+, \bm{k}_\perp) 
&=&
\frac{1}{(\sqrt{2} k^+)^{\frac{1}{2}}}
\left[
\sqrt{2} k^+
- 
\beta m
+ 
\bm{\alpha}_\perp \cdot \bm{k}_\perp
\right]
X_{-\lambda}
.
\notag \\
\end{eqnarray}
The normalization of these spinors is such that
\begin{eqnarray}
\ol u_\lambda (k^+, \bm{k}_\perp) \gamma^+ u_{\lambda'} (q^+, \bm{q}_\perp) 
= 
2 \sqrt{k^+ q^+} \delta_{\lambda, \lambda'}
,\end{eqnarray}
and similarly for the 
$v_\lambda$. 
These spinors have been chosen to agree with those of 
Ref.~\cite{Lepage:1980fj}.

In computing the two-body light-front wavefunction above, 
it is useful to note the product relations
\begin{eqnarray}
\ol u_\lambda (k^+, \bm{k}_\perp) \gamma^+ \left(\rlap \slash k + m\right)
&=& 
2 k^+ 
\ol u_\lambda (k^+, \bm{k}_\perp)
,\notag\\
\left(\rlap \slash k - m \right) \gamma^+ v_\lambda (k^+, \bm{k}_\perp)
&=&
2 k^+ v_\lambda (k^+, \bm{k}_\perp)
,\end{eqnarray}
which hold for 
\emph{any} 
value of 
$k^-$ 
by virtue of the property 
$\left(\gamma^+\right)^2 = 0$. 
Additionally, 
we note that the spinors are eigenvectors of light-front helicity
$\gamma_5$,
namely  
$\gamma_5 \, X_\lambda = \lambda \, X_\lambda$, 
which leads to the relation
%\begin{equation}
$\gamma_5 \, v_\lambda (k^+, \bm{k}_\perp) = - \lambda \, u_{- \lambda} (k^+, \bm{k}_\perp)
$.
%\end{equation} 
In computing the two-body light-front wavefunction, 
we need the spinor product
\begin{eqnarray}
\ol u_\lambda (k^+, \bm{k}_\perp) 
\, \gamma_5 \,
v_{\lambda'} (q^+, \bm{q}_\perp)
&=&
\delta_{\lambda, \lambda'}
\frac{k^+ q_{- \lambda} -  q^+ k_{-\lambda} }{\sqrt{k^+ q^+}}
\notag \\
&&
+ \lambda \, m \, \delta_{\lambda, - \lambda'} 
\frac{k^+ + q^+}{\sqrt{k^+ q^+}}
.\quad \, \, \end{eqnarray}
In the case where  
$q_\mu = (P - k)_\mu$, 
we arrive at the helicity structure in 
Eq.~\eqref{eq:LFWF}.

%%%%%%%%%%%%%%%%%%%%%%%%%%%%%%%%%%%%%%%
\subsection{Normalization}
%%%%%%%%%%%%%%%%%%%%%%%%%%%%%%%%%%%%%%%

To obtain the correct normalization of the two-body light-front wavefunction from the covariant Bethe-Salpeter wavefunction, 
we return to 
Eq.~\eqref{eq:model}
in the case of a point-like vertex, 
$\Gamma(k,P) \equiv \cN$. 
In such a simple model, 
there are no higher Fock components. 
Thus the normalization condition for the two-body wavefunction in this case is fixed by that of the form factor at vanishing momentum transfer. 
We use this fact to establish the normalization convention for the two-body wavefunction in more general models where there are nonvanishing contributions from higher-Fock components. 

In the point-like case, 
the two-body light-front wavefunction obtained from Eq.~\eqref{eq:LFWF2} reads
\begin{eqnarray}
\psi^{(2)}_{\lambda \lambda'} (x, \bm{k}_\perp)
&=&
\cN \,  \frac{k_{-\lambda} \delta_{\lambda,\lambda'} - m \lambda \delta_{\lambda, - \lambda'}}{x(1-x)} 
%\notag \\
%&& \phantom{spa} 
%\times
D_W(x, \bm{k}_\perp; m, m).
\notag \\
\label{eq:pointWF}
\end{eqnarray}
We omit the factor of 
$\theta[x(1-x)]$
for simplicity. 
It remains to show that the normalization condition is 
\begin{equation}
\sum_{\lambda, \lambda'} 
\int \frac{dx \, d\bm{k}_\perp}{2 (2\pi)^3}
\Big|\psi^{(2)}_{\lambda \lambda'} (x, \bm{k}_\perp)\Big|^2
= 
N^{(2)}
,\end{equation}
which is that implicitly used above in defining the two-body quark distribution, 
Eq.~\eqref{eq:q2}. 
In the present point-like model, 
$N^{(2)} = 1$
after suitable regularization;
however, 
in the model of the main text, 
$N^{(2)} < 1$ 
due to higher Fock states.

To show the requirement
$N^{(2)} = 1$, 
we compute the form factor at vanishing momentum transfer in the point-like model. 
As a particular instance of 
Eq.~\eqref{eq:mel}, 
the normalization is fixed by
\begin{equation}
\int_k \Tr \left[ \Psi(k,P) S^{-1}(k-P) \ol \Psi(k,P) \gamma^+ \right] = 2 P^+
\label{eq:pointnorm}
,\end{equation}
where the right-hand side is simply the plus-component of the vector-current matrix element, 
$(P' + P)^\mu F(t)$,
evaluated at vanishing momentum transfer. 
Writing out the elements in Eq.~\eqref{eq:pointnorm}, 
we have the condition
\begin{eqnarray}
\frac{\cN^2}{2 P^+}
\int_k \frac{i N^+ |_{\Delta = 0}}{\cA \, \cC^2} 
=
1
\label{eq:norm}
,\end{eqnarray}
where 
$N^\mu$ 
is the trace appearing in the numerator, 
Eq.~\eqref{eq:numerator}, 
and the scalar propagators, 
$\cA$ 
and
$\cC$,  
are given in 
Eq.~\eqref{eq:props}. 
The plus-component of the trace at vanishing momentum transfer takes the form
\begin{eqnarray}
N^+\big|_{\Delta = 0}
&=& 
4 \left[k^+ ( M^2 - \cA ) - P^+ \cC \right]
\notag \\
&\overset{\text{eff}}{=}&
4 (k^+ M^2 - P^+ \cC)
.\end{eqnarray}
The second equality results from observing that: 
i).~canceling the single 
$\cA$ 
appearing in the denominator leads to light-front 
$k^-$ 
poles always lying in the same half plane;
and,
ii).~while these poles escape to infinity when 
$k^+ \to 0$ 
leading to 
$\delta(k^+)$ 
contributions, 
such contributions are multiplied by 
$k^+$
and accordingly vanish. 
Thus 
$\cA$-reduced light-front 
$k^-$ integrals 
vanish, 
because there are no zero-mode singularities in this model, 
see, e.g., 
Ref.~\cite{Jaus:1999zv}.  
With the trace written as above, 
the 
$k^-$
integral in 
Eq.~\eqref{eq:norm}
can be performed straightforwardly by contour integration. 
This results in
\begin{equation}
\cN^2 
\int \frac{dx \, d \bm{k}_\perp}{2 (2\pi)^3}
\frac{2 (\bm{k}_\perp^2 + m^2)}{x^2 (1-x)^2}
D_W (x, \bm{k}_\perp; m, m )^2=1
.\end{equation}
Comparing this expression with the point-like model's two-body wavefunction in Eq.~\eqref{eq:pointWF}, 
we see that indeed the normalization is fixed to
$N^{(2)} = 1$.

\bibliography{bibly}

%merlin.mbs apsrev4-1.bst 2010-07-25 4.21a (PWD, AO, DPC) hacked
%Control: key (0)
%Control: author (0) dotless jnrlst
%Control: editor formatted (1) identically to author
%Control: production of article title (0) allowed
%Control: page (1) range
%Control: year (0) verbatim
%Control: production of eprint (0) enabled
\begin{thebibliography}{48}%
\makeatletter
\providecommand \@ifxundefined [1]{%
 \@ifx{#1\undefined}
}%
\providecommand \@ifnum [1]{%
 \ifnum #1\expandafter \@firstoftwo
 \else \expandafter \@secondoftwo
 \fi
}%
\providecommand \@ifx [1]{%
 \ifx #1\expandafter \@firstoftwo
 \else \expandafter \@secondoftwo
 \fi
}%
\providecommand \natexlab [1]{#1}%
\providecommand \enquote  [1]{``#1''}%
\providecommand \bibnamefont  [1]{#1}%
\providecommand \bibfnamefont [1]{#1}%
\providecommand \citenamefont [1]{#1}%
\providecommand \href@noop [0]{\@secondoftwo}%
\providecommand \href [0]{\begingroup \@sanitize@url \@href}%
\providecommand \@href[1]{\@@startlink{#1}\@@href}%
\providecommand \@@href[1]{\endgroup#1\@@endlink}%
\providecommand \@sanitize@url [0]{\catcode `\\12\catcode `\$12\catcode
  `\&12\catcode `\#12\catcode `\^12\catcode `\_12\catcode `\%12\relax}%
\providecommand \@@startlink[1]{}%
\providecommand \@@endlink[0]{}%
\providecommand \url  [0]{\begingroup\@sanitize@url \@url }%
\providecommand \@url [1]{\endgroup\@href {#1}{\urlprefix }}%
\providecommand \urlprefix  [0]{URL }%
\providecommand \Eprint [0]{\href }%
\providecommand \doibase [0]{http://dx.doi.org/}%
\providecommand \selectlanguage [0]{\@gobble}%
\providecommand \bibinfo  [0]{\@secondoftwo}%
\providecommand \bibfield  [0]{\@secondoftwo}%
\providecommand \translation [1]{[#1]}%
\providecommand \BibitemOpen [0]{}%
\providecommand \bibitemStop [0]{}%
\providecommand \bibitemNoStop [0]{.\EOS\space}%
\providecommand \EOS [0]{\spacefactor3000\relax}%
\providecommand \BibitemShut  [1]{\csname bibitem#1\endcsname}%
\let\auto@bib@innerbib\@empty
%</preamble>
\bibitem [{\citenamefont {{D.~M\"uller, D.~Robaschik, B.~Geyer, F.-M.~Dittes,
  J.~Ho\v{r}ej\v{s}i}}(1994)}]{Mueller:1998fv}%
  \BibitemOpen
  \bibfield  {author} {\bibinfo {author} {\bibnamefont {{D.~M\"uller,
  D.~Robaschik, B.~Geyer, F.-M.~Dittes, J.~Ho\v{r}ej\v{s}i}}},\ }\bibfield
  {title} {\enquote {\bibinfo {title} {{Wave functions, evolution equations and
  evolution kernels from light ray operators of QCD}},}\ }\href {\doibase
  10.1002/prop.2190420202} {\bibfield  {journal} {\bibinfo  {journal} {Fortsch.
  Phys.}\ }\textbf {\bibinfo {volume} {42}},\ \bibinfo {pages} {101--141}
  (\bibinfo {year} {1994})},\ \Eprint {http://arxiv.org/abs/hep-ph/9812448}
  {arXiv:hep-ph/9812448 [hep-ph]} \BibitemShut {NoStop}%
%%CITATION = HEP-PH/9812448;%%
\bibitem [{\citenamefont {Ji}(1997{\natexlab{a}})}]{Ji:1996ek}%
  \BibitemOpen
  \bibfield  {author} {\bibinfo {author} {\bibfnamefont {X.-D.}\ \bibnamefont
  {Ji}},\ }\bibfield  {title} {\enquote {\bibinfo {title} {{Gauge-invariant
  decomposition of nucleon spin}},}\ }\href {\doibase
  10.1103/PhysRevLett.78.610} {\bibfield  {journal} {\bibinfo  {journal} {Phys.
  Rev. Lett.}\ }\textbf {\bibinfo {volume} {78}},\ \bibinfo {pages} {610--613}
  (\bibinfo {year} {1997}{\natexlab{a}})},\ \Eprint
  {http://arxiv.org/abs/hep-ph/9603249} {arXiv:hep-ph/9603249 [hep-ph]}
  \BibitemShut {NoStop}%
%%CITATION = HEP-PH/9603249;%%
\bibitem [{\citenamefont {Ji}(1997{\natexlab{b}})}]{Ji:1996nm}%
  \BibitemOpen
  \bibfield  {author} {\bibinfo {author} {\bibfnamefont {X.-D.}\ \bibnamefont
  {Ji}},\ }\bibfield  {title} {\enquote {\bibinfo {title} {{Deeply virtual
  Compton scattering}},}\ }\href {\doibase 10.1103/PhysRevD.55.7114} {\bibfield
   {journal} {\bibinfo  {journal} {Phys. Rev.}\ }\textbf {\bibinfo {volume}
  {D55}},\ \bibinfo {pages} {7114--7125} (\bibinfo {year}
  {1997}{\natexlab{b}})},\ \Eprint {http://arxiv.org/abs/hep-ph/9609381}
  {arXiv:hep-ph/9609381 [hep-ph]} \BibitemShut {NoStop}%
%%CITATION = HEP-PH/9609381;%%
\bibitem [{\citenamefont {Radyushkin}(1996{\natexlab{a}})}]{Radyushkin:1996nd}%
  \BibitemOpen
  \bibfield  {author} {\bibinfo {author} {\bibfnamefont {A.~V.}\ \bibnamefont
  {Radyushkin}},\ }\bibfield  {title} {\enquote {\bibinfo {title} {{Scaling
  limit of deeply virtual Compton scattering}},}\ }\href {\doibase
  10.1016/0370-2693(96)00528-X} {\bibfield  {journal} {\bibinfo  {journal}
  {Phys. Lett.}\ }\textbf {\bibinfo {volume} {B380}},\ \bibinfo {pages}
  {417--425} (\bibinfo {year} {1996}{\natexlab{a}})},\ \Eprint
  {http://arxiv.org/abs/hep-ph/9604317} {arXiv:hep-ph/9604317 [hep-ph]}
  \BibitemShut {NoStop}%
%%CITATION = HEP-PH/9604317;%%
\bibitem [{\citenamefont {Radyushkin}(1996{\natexlab{b}})}]{Radyushkin:1996ru}%
  \BibitemOpen
  \bibfield  {author} {\bibinfo {author} {\bibfnamefont {A.~V.}\ \bibnamefont
  {Radyushkin}},\ }\bibfield  {title} {\enquote {\bibinfo {title} {{Asymmetric
  gluon distributions and hard diffractive electroproduction}},}\ }\href
  {\doibase 10.1016/0370-2693(96)00844-1} {\bibfield  {journal} {\bibinfo
  {journal} {Phys. Lett.}\ }\textbf {\bibinfo {volume} {B385}},\ \bibinfo
  {pages} {333--342} (\bibinfo {year} {1996}{\natexlab{b}})},\ \Eprint
  {http://arxiv.org/abs/hep-ph/9605431} {arXiv:hep-ph/9605431 [hep-ph]}
  \BibitemShut {NoStop}%
%%CITATION = HEP-PH/9605431;%%
\bibitem [{\citenamefont {Burkardt}(2000)}]{Burkardt:2000za}%
  \BibitemOpen
  \bibfield  {author} {\bibinfo {author} {\bibfnamefont {M.}~\bibnamefont
  {Burkardt}},\ }\bibfield  {title} {\enquote {\bibinfo {title} {{Impact
  parameter dependent parton distributions and off forward parton distributions
  for $\zeta \to 0$}},}\ }\href {\doibase 10.1103/PhysRevD.62.071503,
  10.1103/PhysRevD.66.119903} {\bibfield  {journal} {\bibinfo  {journal} {Phys.
  Rev.}\ }\textbf {\bibinfo {volume} {D62}},\ \bibinfo {pages} {071503}
  (\bibinfo {year} {2000})},\ \bibinfo {note} {[Erratum: Phys.
  Rev.D66,119903(2002)]},\ \Eprint {http://arxiv.org/abs/hep-ph/0005108}
  {arXiv:hep-ph/0005108 [hep-ph]} \BibitemShut {NoStop}%
%%CITATION = HEP-PH/0005108;%%
\bibitem [{\citenamefont {Burkardt}(2003)}]{Burkardt:2002hr}%
  \BibitemOpen
  \bibfield  {author} {\bibinfo {author} {\bibfnamefont {M.}~\bibnamefont
  {Burkardt}},\ }\bibfield  {title} {\enquote {\bibinfo {title} {{Impact
  parameter space interpretation for generalized parton distributions}},}\
  }\href {\doibase 10.1142/S0217751X03012370} {\bibfield  {journal} {\bibinfo
  {journal} {Int. J. Mod. Phys.}\ }\textbf {\bibinfo {volume} {A18}},\ \bibinfo
  {pages} {173--208} (\bibinfo {year} {2003})},\ \Eprint
  {http://arxiv.org/abs/hep-ph/0207047} {arXiv:hep-ph/0207047 [hep-ph]}
  \BibitemShut {NoStop}%
%%CITATION = HEP-PH/0207047;%%
\bibitem [{\citenamefont {Diehl}(2002)}]{Diehl:2002he}%
  \BibitemOpen
  \bibfield  {author} {\bibinfo {author} {\bibfnamefont {M.}~\bibnamefont
  {Diehl}},\ }\bibfield  {title} {\enquote {\bibinfo {title} {{Generalized
  parton distributions in impact parameter space}},}\ }\href {\doibase
  10.1007/s10052-002-1016-9} {\bibfield  {journal} {\bibinfo  {journal} {Eur.
  Phys. J.}\ }\textbf {\bibinfo {volume} {C25}},\ \bibinfo {pages} {223--232}
  (\bibinfo {year} {2002})},\ \bibinfo {note} {[Erratum: Eur. Phys.
  J.C31,277(2003)]},\ \Eprint {http://arxiv.org/abs/hep-ph/0205208}
  {arXiv:hep-ph/0205208 [hep-ph]} \BibitemShut {NoStop}%
%%CITATION = HEP-PH/0205208;%%
\bibitem [{\citenamefont {Goeke}\ \emph {et~al.}(2001)\citenamefont {Goeke},
  \citenamefont {Polyakov},\ and\ \citenamefont
  {Vanderhaeghen}}]{Goeke:2001tz}%
  \BibitemOpen
  \bibfield  {author} {\bibinfo {author} {\bibfnamefont {K.}~\bibnamefont
  {Goeke}}, \bibinfo {author} {\bibfnamefont {M.~V.}\ \bibnamefont {Polyakov}},
  \ and\ \bibinfo {author} {\bibfnamefont {M.}~\bibnamefont {Vanderhaeghen}},\
  }\bibfield  {title} {\enquote {\bibinfo {title} {{Hard exclusive reactions
  and the structure of hadrons}},}\ }\href {\doibase
  10.1016/S0146-6410(01)00158-2} {\bibfield  {journal} {\bibinfo  {journal}
  {Prog. Part. Nucl. Phys.}\ }\textbf {\bibinfo {volume} {47}},\ \bibinfo
  {pages} {401--515} (\bibinfo {year} {2001})},\ \Eprint
  {http://arxiv.org/abs/hep-ph/0106012} {arXiv:hep-ph/0106012 [hep-ph]}
  \BibitemShut {NoStop}%
%%CITATION = HEP-PH/0106012;%%
\bibitem [{\citenamefont {Diehl}(2003)}]{Diehl:2003ny}%
  \BibitemOpen
  \bibfield  {author} {\bibinfo {author} {\bibfnamefont {M.}~\bibnamefont
  {Diehl}},\ }\bibfield  {title} {\enquote {\bibinfo {title} {{Generalized
  parton distributions}},}\ }\href {\doibase 10.1016/j.physrep.2003.08.002,
  10.3204/DESY-THESIS-2003-018} {\bibfield  {journal} {\bibinfo  {journal}
  {Phys. Rept.}\ }\textbf {\bibinfo {volume} {388}},\ \bibinfo {pages}
  {41--277} (\bibinfo {year} {2003})},\ \Eprint
  {http://arxiv.org/abs/hep-ph/0307382} {arXiv:hep-ph/0307382 [hep-ph]}
  \BibitemShut {NoStop}%
%%CITATION = HEP-PH/0307382;%%
\bibitem [{\citenamefont {Belitsky}\ and\ \citenamefont
  {Radyushkin}(2005)}]{Belitsky:2005qn}%
  \BibitemOpen
  \bibfield  {author} {\bibinfo {author} {\bibfnamefont {A.~V.}\ \bibnamefont
  {Belitsky}}\ and\ \bibinfo {author} {\bibfnamefont {A.~V.}\ \bibnamefont
  {Radyushkin}},\ }\bibfield  {title} {\enquote {\bibinfo {title} {{Unraveling
  hadron structure with generalized parton distributions}},}\ }\href {\doibase
  10.1016/j.physrep.2005.06.002} {\bibfield  {journal} {\bibinfo  {journal}
  {Phys. Rept.}\ }\textbf {\bibinfo {volume} {418}},\ \bibinfo {pages} {1--387}
  (\bibinfo {year} {2005})},\ \Eprint {http://arxiv.org/abs/hep-ph/0504030}
  {arXiv:hep-ph/0504030 [hep-ph]} \BibitemShut {NoStop}%
%%CITATION = HEP-PH/0504030;%%
\bibitem [{\citenamefont {Boffi}\ and\ \citenamefont
  {Pasquini}(2007)}]{Boffi:2007yc}%
  \BibitemOpen
  \bibfield  {author} {\bibinfo {author} {\bibfnamefont {S.}~\bibnamefont
  {Boffi}}\ and\ \bibinfo {author} {\bibfnamefont {B.}~\bibnamefont
  {Pasquini}},\ }\bibfield  {title} {\enquote {\bibinfo {title} {{Generalized
  parton distributions and the structure of the nucleon}},}\ }\href {\doibase
  10.1393/ncr/i2007-10025-7} {\bibfield  {journal} {\bibinfo  {journal} {Riv.
  Nuovo Cim.}\ }\textbf {\bibinfo {volume} {30}},\ \bibinfo {pages} {387}
  (\bibinfo {year} {2007})},\ \Eprint {http://arxiv.org/abs/0711.2625}
  {arXiv:0711.2625 [hep-ph]} \BibitemShut {NoStop}%
%%CITATION = ARXIV:0711.2625;%%
\bibitem [{\citenamefont {Martin}\ and\ \citenamefont
  {Ryskin}(1998)}]{Martin:1997wy}%
  \BibitemOpen
  \bibfield  {author} {\bibinfo {author} {\bibfnamefont {A.~D.}\ \bibnamefont
  {Martin}}\ and\ \bibinfo {author} {\bibfnamefont {M.~G.}\ \bibnamefont
  {Ryskin}},\ }\bibfield  {title} {\enquote {\bibinfo {title} {{The effect of
  off diagonal parton distributions in diffractive vector meson
  electroproduction}},}\ }\href {\doibase 10.1103/PhysRevD.57.6692} {\bibfield
  {journal} {\bibinfo  {journal} {Phys. Rev.}\ }\textbf {\bibinfo {volume}
  {D57}},\ \bibinfo {pages} {6692--6700} (\bibinfo {year} {1998})},\ \Eprint
  {http://arxiv.org/abs/hep-ph/9711371} {arXiv:hep-ph/9711371 [hep-ph]}
  \BibitemShut {NoStop}%
%%CITATION = HEP-PH/9711371;%%
\bibitem [{\citenamefont {Ji}(1998)}]{Ji:1998pc}%
  \BibitemOpen
  \bibfield  {author} {\bibinfo {author} {\bibfnamefont {X.-D.}\ \bibnamefont
  {Ji}},\ }\bibfield  {title} {\enquote {\bibinfo {title} {{Off forward parton
  distributions}},}\ }\href {\doibase 10.1088/0954-3899/24/7/002} {\bibfield
  {journal} {\bibinfo  {journal} {J. Phys.}\ }\textbf {\bibinfo {volume}
  {G24}},\ \bibinfo {pages} {1181--1205} (\bibinfo {year} {1998})},\ \Eprint
  {http://arxiv.org/abs/hep-ph/9807358} {arXiv:hep-ph/9807358 [hep-ph]}
  \BibitemShut {NoStop}%
%%CITATION = HEP-PH/9807358;%%
\bibitem [{\citenamefont {Pire}\ \emph {et~al.}(1999)\citenamefont {Pire},
  \citenamefont {Soffer},\ and\ \citenamefont {Teryaev}}]{Pire:1998nw}%
  \BibitemOpen
  \bibfield  {author} {\bibinfo {author} {\bibfnamefont {B.}~\bibnamefont
  {Pire}}, \bibinfo {author} {\bibfnamefont {J.}~\bibnamefont {Soffer}}, \ and\
  \bibinfo {author} {\bibfnamefont {O.}~\bibnamefont {Teryaev}},\ }\bibfield
  {title} {\enquote {\bibinfo {title} {{Positivity constraints for off-forward
  parton distributions}},}\ }\href {\doibase 10.1007/s100529901063} {\bibfield
  {journal} {\bibinfo  {journal} {Eur. Phys. J.}\ }\textbf {\bibinfo {volume}
  {C8}},\ \bibinfo {pages} {103--106} (\bibinfo {year} {1999})},\ \Eprint
  {http://arxiv.org/abs/hep-ph/9804284} {arXiv:hep-ph/9804284 [hep-ph]}
  \BibitemShut {NoStop}%
%%CITATION = HEP-PH/9804284;%%
\bibitem [{\citenamefont {Radyushkin}(1999)}]{Radyushkin:1998es}%
  \BibitemOpen
  \bibfield  {author} {\bibinfo {author} {\bibfnamefont {A.~V.}\ \bibnamefont
  {Radyushkin}},\ }\bibfield  {title} {\enquote {\bibinfo {title} {{Double
  distributions and evolution equations}},}\ }\href {\doibase
  10.1103/PhysRevD.59.014030} {\bibfield  {journal} {\bibinfo  {journal} {Phys.
  Rev.}\ }\textbf {\bibinfo {volume} {D59}},\ \bibinfo {pages} {014030}
  (\bibinfo {year} {1999})},\ \Eprint {http://arxiv.org/abs/hep-ph/9805342}
  {arXiv:hep-ph/9805342 [hep-ph]} \BibitemShut {NoStop}%
%%CITATION = HEP-PH/9805342;%%
\bibitem [{\citenamefont {Pobylitsa}(2002{\natexlab{a}})}]{Pobylitsa:2001nt}%
  \BibitemOpen
  \bibfield  {author} {\bibinfo {author} {\bibfnamefont {P.~V.}\ \bibnamefont
  {Pobylitsa}},\ }\bibfield  {title} {\enquote {\bibinfo {title} {{Inequalities
  for generalized parton distributions $H$ and $E$}},}\ }\href {\doibase
  10.1103/PhysRevD.65.077504} {\bibfield  {journal} {\bibinfo  {journal} {Phys.
  Rev.}\ }\textbf {\bibinfo {volume} {D65}},\ \bibinfo {pages} {077504}
  (\bibinfo {year} {2002}{\natexlab{a}})},\ \Eprint
  {http://arxiv.org/abs/hep-ph/0112322} {arXiv:hep-ph/0112322 [hep-ph]}
  \BibitemShut {NoStop}%
%%CITATION = HEP-PH/0112322;%%
\bibitem [{\citenamefont {Pobylitsa}(2002{\natexlab{b}})}]{Pobylitsa:2002gw}%
  \BibitemOpen
  \bibfield  {author} {\bibinfo {author} {\bibfnamefont {P.~V.}\ \bibnamefont
  {Pobylitsa}},\ }\bibfield  {title} {\enquote {\bibinfo {title}
  {{Disentangling positivity constraints for generalized parton
  distributions}},}\ }\href {\doibase 10.1103/PhysRevD.65.114015} {\bibfield
  {journal} {\bibinfo  {journal} {Phys. Rev.}\ }\textbf {\bibinfo {volume}
  {D65}},\ \bibinfo {pages} {114015} (\bibinfo {year} {2002}{\natexlab{b}})},\
  \Eprint {http://arxiv.org/abs/hep-ph/0201030} {arXiv:hep-ph/0201030 [hep-ph]}
  \BibitemShut {NoStop}%
%%CITATION = HEP-PH/0201030;%%
\bibitem [{\citenamefont {Brodsky}\ \emph {et~al.}(2001)\citenamefont
  {Brodsky}, \citenamefont {Diehl},\ and\ \citenamefont
  {Hwang}}]{Brodsky:2000xy}%
  \BibitemOpen
  \bibfield  {author} {\bibinfo {author} {\bibfnamefont {S.~J.}\ \bibnamefont
  {Brodsky}}, \bibinfo {author} {\bibfnamefont {M.}~\bibnamefont {Diehl}}, \
  and\ \bibinfo {author} {\bibfnamefont {D.-S.}\ \bibnamefont {Hwang}},\
  }\bibfield  {title} {\enquote {\bibinfo {title} {{Light cone wave function
  representation of deeply virtual Compton scattering}},}\ }\href {\doibase
  10.1016/S0550-3213(00)00695-7} {\bibfield  {journal} {\bibinfo  {journal}
  {Nucl. Phys.}\ }\textbf {\bibinfo {volume} {B596}},\ \bibinfo {pages}
  {99--124} (\bibinfo {year} {2001})},\ \Eprint
  {http://arxiv.org/abs/hep-ph/0009254} {arXiv:hep-ph/0009254 [hep-ph]}
  \BibitemShut {NoStop}%
%%CITATION = HEP-PH/0009254;%%
\bibitem [{\citenamefont {Diehl}\ \emph {et~al.}(2001)\citenamefont {Diehl},
  \citenamefont {Feldmann}, \citenamefont {Jakob},\ and\ \citenamefont
  {Kroll}}]{Diehl:2000xz}%
  \BibitemOpen
  \bibfield  {author} {\bibinfo {author} {\bibfnamefont {M.}~\bibnamefont
  {Diehl}}, \bibinfo {author} {\bibfnamefont {T.}~\bibnamefont {Feldmann}},
  \bibinfo {author} {\bibfnamefont {R.}~\bibnamefont {Jakob}}, \ and\ \bibinfo
  {author} {\bibfnamefont {P.}~\bibnamefont {Kroll}},\ }\bibfield  {title}
  {\enquote {\bibinfo {title} {{The overlap representation of skewed quark and
  gluon distributions}},}\ }\href {\doibase 10.1016/S0550-3213(00)00684-2,
  10.1016/S0550-3213(01)00183-3} {\bibfield  {journal} {\bibinfo  {journal}
  {Nucl. Phys.}\ }\textbf {\bibinfo {volume} {B596}},\ \bibinfo {pages}
  {33--65} (\bibinfo {year} {2001})},\ \bibinfo {note} {[Erratum: Nucl.
  Phys.B605,647(2001)]},\ \Eprint {http://arxiv.org/abs/hep-ph/0009255}
  {arXiv:hep-ph/0009255 [hep-ph]} \BibitemShut {NoStop}%
%%CITATION = HEP-PH/0009255;%%
\bibitem [{\citenamefont {Radyushkin}(1997)}]{Radyushkin:1997ki}%
  \BibitemOpen
  \bibfield  {author} {\bibinfo {author} {\bibfnamefont {A.~V.}\ \bibnamefont
  {Radyushkin}},\ }\bibfield  {title} {\enquote {\bibinfo {title} {{Nonforward
  parton distributions}},}\ }\href {\doibase 10.1103/PhysRevD.56.5524}
  {\bibfield  {journal} {\bibinfo  {journal} {Phys. Rev.}\ }\textbf {\bibinfo
  {volume} {D56}},\ \bibinfo {pages} {5524--5557} (\bibinfo {year} {1997})},\
  \Eprint {http://arxiv.org/abs/hep-ph/9704207} {arXiv:hep-ph/9704207 [hep-ph]}
  \BibitemShut {NoStop}%
%%CITATION = HEP-PH/9704207;%%
\bibitem [{\citenamefont {Fanelli}\ \emph {et~al.}(2016)\citenamefont
  {Fanelli}, \citenamefont {Pace}, \citenamefont {Romanelli}, \citenamefont
  {Salm\`e},\ and\ \citenamefont {Salmistraro}}]{Fanelli:2016aqc}%
  \BibitemOpen
  \bibfield  {author} {\bibinfo {author} {\bibfnamefont {C.}~\bibnamefont
  {Fanelli}}, \bibinfo {author} {\bibfnamefont {E.}~\bibnamefont {Pace}},
  \bibinfo {author} {\bibfnamefont {G.}~\bibnamefont {Romanelli}}, \bibinfo
  {author} {\bibfnamefont {G.}~\bibnamefont {Salm\`e}}, \ and\ \bibinfo
  {author} {\bibfnamefont {M.}~\bibnamefont {Salmistraro}},\ }\bibfield
  {title} {\enquote {\bibinfo {title} {Pion generalized parton distributions
  within a fully covariant constituent quark model},}\ }\href {\doibase
  10.1140/epjc/s10052-016-4101-1} {\bibfield  {journal} {\bibinfo  {journal}
  {Eur. Phys. J.}\ }\textbf {\bibinfo {volume} {C76}},\ \bibinfo {pages} {253}
  (\bibinfo {year} {2016})},\ \Eprint {http://arxiv.org/abs/1603.04598}
  {arXiv:1603.04598 [hep-ph]} \BibitemShut {NoStop}%
%%CITATION = ARXIV:1603.04598;%%
\bibitem [{\citenamefont {Theussl}\ \emph {et~al.}(2004)\citenamefont
  {Theussl}, \citenamefont {Noguera},\ and\ \citenamefont
  {Vento}}]{Theussl:2002xp}%
  \BibitemOpen
  \bibfield  {author} {\bibinfo {author} {\bibfnamefont {L.}~\bibnamefont
  {Theussl}}, \bibinfo {author} {\bibfnamefont {S.}~\bibnamefont {Noguera}}, \
  and\ \bibinfo {author} {\bibfnamefont {V.}~\bibnamefont {Vento}},\ }\bibfield
   {title} {\enquote {\bibinfo {title} {{Generalized parton distributions of
  the pion in a Bethe-Salpeter approach}},}\ }\href {\doibase
  10.1140/epja/i2003-10174-3} {\bibfield  {journal} {\bibinfo  {journal} {Eur.
  Phys. J.}\ }\textbf {\bibinfo {volume} {A20}},\ \bibinfo {pages} {483--498}
  (\bibinfo {year} {2004})},\ \Eprint {http://arxiv.org/abs/nucl-th/0211036}
  {arXiv:nucl-th/0211036 [nucl-th]} \BibitemShut {NoStop}%
%%CITATION = NUCL-TH/0211036;%%
\bibitem [{\citenamefont {Broniowski}\ \emph {et~al.}(2008)\citenamefont
  {Broniowski}, \citenamefont {Ruiz~Arriola},\ and\ \citenamefont
  {Golec-Biernat}}]{Broniowski:2007si}%
  \BibitemOpen
  \bibfield  {author} {\bibinfo {author} {\bibfnamefont {W.}~\bibnamefont
  {Broniowski}}, \bibinfo {author} {\bibfnamefont {E.}~\bibnamefont
  {Ruiz~Arriola}}, \ and\ \bibinfo {author} {\bibfnamefont {K.}~\bibnamefont
  {Golec-Biernat}},\ }\bibfield  {title} {\enquote {\bibinfo {title}
  {{Generalized parton distributions of the pion in chiral quark models and
  their QCD evolution}},}\ }\href {\doibase 10.1103/PhysRevD.77.034023}
  {\bibfield  {journal} {\bibinfo  {journal} {Phys. Rev.}\ }\textbf {\bibinfo
  {volume} {D77}},\ \bibinfo {pages} {034023} (\bibinfo {year} {2008})},\
  \Eprint {http://arxiv.org/abs/0712.1012} {arXiv:0712.1012 [hep-ph]}
  \BibitemShut {NoStop}%
%%CITATION = ARXIV:0712.1012;%%
\bibitem [{\citenamefont {Davidson}\ and\ \citenamefont
  {Ruiz~Arriola}(1995)}]{Davidson:1994uv}%
  \BibitemOpen
  \bibfield  {author} {\bibinfo {author} {\bibfnamefont {R.~M.}\ \bibnamefont
  {Davidson}}\ and\ \bibinfo {author} {\bibfnamefont {E.}~\bibnamefont
  {Ruiz~Arriola}},\ }\bibfield  {title} {\enquote {\bibinfo {title} {{Structure
  functions of pseudoscalar mesons in the $SU(3)$ NJL model}},}\ }\href
  {\doibase 10.1016/0370-2693(95)00091-X} {\bibfield  {journal} {\bibinfo
  {journal} {Phys. Lett.}\ }\textbf {\bibinfo {volume} {B348}},\ \bibinfo
  {pages} {163--169} (\bibinfo {year} {1995})}\BibitemShut {NoStop}%
%%CITATION = PHLTA,B348,163;%%
\bibitem [{\citenamefont {Pobylitsa}(2003)}]{Pobylitsa:2002vw}%
  \BibitemOpen
  \bibfield  {author} {\bibinfo {author} {\bibfnamefont {P.~V.}\ \bibnamefont
  {Pobylitsa}},\ }\bibfield  {title} {\enquote {\bibinfo {title} {{Integral
  representations for nonperturbative GPDs in terms of perturbative
  diagrams}},}\ }\href {\doibase 10.1103/PhysRevD.67.094012} {\bibfield
  {journal} {\bibinfo  {journal} {Phys. Rev.}\ }\textbf {\bibinfo {volume}
  {D67}},\ \bibinfo {pages} {094012} (\bibinfo {year} {2003})},\ \Eprint
  {http://arxiv.org/abs/hep-ph/0210238} {arXiv:hep-ph/0210238 [hep-ph]}
  \BibitemShut {NoStop}%
%%CITATION = HEP-PH/0210238;%%
\bibitem [{\citenamefont {Karmanov}\ and\ \citenamefont
  {Carbonell}(2006)}]{Karmanov:2005nv}%
  \BibitemOpen
  \bibfield  {author} {\bibinfo {author} {\bibfnamefont {V.~A.}\ \bibnamefont
  {Karmanov}}\ and\ \bibinfo {author} {\bibfnamefont {J.}~\bibnamefont
  {Carbonell}},\ }\bibfield  {title} {\enquote {\bibinfo {title} {{Solving
  Bethe-Salpeter equation in Minkowski space}},}\ }\href {\doibase
  10.1140/epja/i2005-10193-0} {\bibfield  {journal} {\bibinfo  {journal} {Eur.
  Phys. J.}\ }\textbf {\bibinfo {volume} {A27}},\ \bibinfo {pages} {1--9}
  (\bibinfo {year} {2006})},\ \Eprint {http://arxiv.org/abs/hep-th/0505261}
  {arXiv:hep-th/0505261 [hep-th]} \BibitemShut {NoStop}%
%%CITATION = HEP-TH/0505261;%%
\bibitem [{\citenamefont {Carbonell}\ \emph {et~al.}(2009)\citenamefont
  {Carbonell}, \citenamefont {Karmanov},\ and\ \citenamefont
  {Mangin-Brinet}}]{Carbonell:2008tz}%
  \BibitemOpen
  \bibfield  {author} {\bibinfo {author} {\bibfnamefont {J.}~\bibnamefont
  {Carbonell}}, \bibinfo {author} {\bibfnamefont {V.~A.}\ \bibnamefont
  {Karmanov}}, \ and\ \bibinfo {author} {\bibfnamefont {M.}~\bibnamefont
  {Mangin-Brinet}},\ }\bibfield  {title} {\enquote {\bibinfo {title}
  {{Electromagnetic form factor via Bethe-Salpeter amplitude in Minkowski
  space}},}\ }\href {\doibase 10.1140/epja/i2008-10690-6} {\bibfield  {journal}
  {\bibinfo  {journal} {Eur. Phys. J.}\ }\textbf {\bibinfo {volume} {A39}},\
  \bibinfo {pages} {53--60} (\bibinfo {year} {2009})},\ \Eprint
  {http://arxiv.org/abs/0809.3678} {arXiv:0809.3678 [hep-ph]} \BibitemShut
  {NoStop}%
%%CITATION = ARXIV:0809.3678;%%
\bibitem [{\citenamefont {Carbonell}\ \emph {et~al.}(2017)\citenamefont
  {Carbonell}, \citenamefont {Frederico},\ and\ \citenamefont
  {Karmanov}}]{Carbonell:2017kqa}%
  \BibitemOpen
  \bibfield  {author} {\bibinfo {author} {\bibfnamefont {J.}~\bibnamefont
  {Carbonell}}, \bibinfo {author} {\bibfnamefont {T.}~\bibnamefont
  {Frederico}}, \ and\ \bibinfo {author} {\bibfnamefont {V.~A.}\ \bibnamefont
  {Karmanov}},\ }\bibfield  {title} {\enquote {\bibinfo {title} {{Bound state
  equation for the Nakanishi weight function}},}\ }\href {\doibase
  10.1016/j.physletb.2017.04.016} {\bibfield  {journal} {\bibinfo  {journal}
  {Phys. Lett.}\ }\textbf {\bibinfo {volume} {B769}},\ \bibinfo {pages}
  {418--423} (\bibinfo {year} {2017})},\ \Eprint
  {http://arxiv.org/abs/1704.04160} {arXiv:1704.04160 [hep-ph]} \BibitemShut
  {NoStop}%
%%CITATION = ARXIV:1704.04160;%%
\bibitem [{\citenamefont {Nakanishi}(1963)}]{Nakanishi:1963zz}%
  \BibitemOpen
  \bibfield  {author} {\bibinfo {author} {\bibfnamefont {N.}~\bibnamefont
  {Nakanishi}},\ }\bibfield  {title} {\enquote {\bibinfo {title} {{Partial-wave
  Bethe-Salpeter equation}},}\ }\href {\doibase 10.1103/PhysRev.130.1230}
  {\bibfield  {journal} {\bibinfo  {journal} {Phys. Rev.}\ }\textbf {\bibinfo
  {volume} {130}},\ \bibinfo {pages} {1230--1235} (\bibinfo {year}
  {1963})}\BibitemShut {NoStop}%
%%CITATION = PHRVA,130,1230;%%
\bibitem [{\citenamefont {Nakanishi}(1969)}]{Nakanishi:1969ph}%
  \BibitemOpen
  \bibfield  {author} {\bibinfo {author} {\bibfnamefont {N.}~\bibnamefont
  {Nakanishi}},\ }\bibfield  {title} {\enquote {\bibinfo {title} {{A general
  survey of the theory of the Bethe-Salpeter equation}},}\ }\href {\doibase
  10.1143/PTPS.43.1} {\bibfield  {journal} {\bibinfo  {journal} {Prog. Theor.
  Phys. Suppl.}\ }\textbf {\bibinfo {volume} {43}},\ \bibinfo {pages} {1--81}
  (\bibinfo {year} {1969})}\BibitemShut {NoStop}%
%%CITATION = PTPSA,43,1;%%
\bibitem [{\citenamefont {Nguyen}\ \emph {et~al.}(2011)\citenamefont {Nguyen},
  \citenamefont {Bashir}, \citenamefont {Roberts},\ and\ \citenamefont
  {Tandy}}]{Nguyen:2011jy}%
  \BibitemOpen
  \bibfield  {author} {\bibinfo {author} {\bibfnamefont {T.}~\bibnamefont
  {Nguyen}}, \bibinfo {author} {\bibfnamefont {A.}~\bibnamefont {Bashir}},
  \bibinfo {author} {\bibfnamefont {C.~D.}\ \bibnamefont {Roberts}}, \ and\
  \bibinfo {author} {\bibfnamefont {P.~C.}\ \bibnamefont {Tandy}},\ }\bibfield
  {title} {\enquote {\bibinfo {title} {{Pion and kaon valence-quark parton
  distribution functions}},}\ }\href {\doibase 10.1103/PhysRevC.83.062201}
  {\bibfield  {journal} {\bibinfo  {journal} {Phys. Rev.}\ }\textbf {\bibinfo
  {volume} {C83}},\ \bibinfo {pages} {062201} (\bibinfo {year} {2011})},\
  \Eprint {http://arxiv.org/abs/1102.2448} {arXiv:1102.2448 [nucl-th]}
  \BibitemShut {NoStop}%
%%CITATION = ARXIV:1102.2448;%%
\bibitem [{\citenamefont {Mezrag}\ \emph {et~al.}(2015)\citenamefont {Mezrag},
  \citenamefont {Chang}, \citenamefont {Moutarde}, \citenamefont {Roberts},
  \citenamefont {Rodr'guez-Quintero}, \citenamefont {Sabati\'e},\ and\
  \citenamefont {Schmidt}}]{Mezrag:2014jka}%
  \BibitemOpen
  \bibfield  {author} {\bibinfo {author} {\bibfnamefont {C.}~\bibnamefont
  {Mezrag}}, \bibinfo {author} {\bibfnamefont {L.}~\bibnamefont {Chang}},
  \bibinfo {author} {\bibfnamefont {H.}~\bibnamefont {Moutarde}}, \bibinfo
  {author} {\bibfnamefont {C.~D.}\ \bibnamefont {Roberts}}, \bibinfo {author}
  {\bibfnamefont {J.}~\bibnamefont {Rodr'guez-Quintero}}, \bibinfo {author}
  {\bibfnamefont {F.}~\bibnamefont {Sabati\'e}}, \ and\ \bibinfo {author}
  {\bibfnamefont {S.~M.}\ \bibnamefont {Schmidt}},\ }\bibfield  {title}
  {\enquote {\bibinfo {title} {{Sketching the pion's valence-quark generalised
  parton distribution}},}\ }\href {\doibase 10.1016/j.physletb.2014.12.027}
  {\bibfield  {journal} {\bibinfo  {journal} {Phys. Lett.}\ }\textbf {\bibinfo
  {volume} {B741}},\ \bibinfo {pages} {190--196} (\bibinfo {year} {2015})},\
  \Eprint {http://arxiv.org/abs/1411.6634} {arXiv:1411.6634 [nucl-th]}
  \BibitemShut {NoStop}%
%%CITATION = ARXIV:1411.6634;%%
\bibitem [{\citenamefont {Teryaev}(2001)}]{Teryaev:2001qm}%
  \BibitemOpen
  \bibfield  {author} {\bibinfo {author} {\bibfnamefont {O.~V.}\ \bibnamefont
  {Teryaev}},\ }\bibfield  {title} {\enquote {\bibinfo {title} {{Crossing and
  radon tomography for generalized parton distributions}},}\ }\href {\doibase
  10.1016/S0370-2693(01)00564-0} {\bibfield  {journal} {\bibinfo  {journal}
  {Phys. Lett.}\ }\textbf {\bibinfo {volume} {B510}},\ \bibinfo {pages}
  {125--132} (\bibinfo {year} {2001})},\ \Eprint
  {http://arxiv.org/abs/hep-ph/0102303} {arXiv:hep-ph/0102303 [hep-ph]}
  \BibitemShut {NoStop}%
%%CITATION = HEP-PH/0102303;%%
\bibitem [{\citenamefont {Polyakov}\ and\ \citenamefont
  {Weiss}(1999)}]{Polyakov:1999gs}%
  \BibitemOpen
  \bibfield  {author} {\bibinfo {author} {\bibfnamefont {M.~V.}\ \bibnamefont
  {Polyakov}}\ and\ \bibinfo {author} {\bibfnamefont {C.}~\bibnamefont
  {Weiss}},\ }\bibfield  {title} {\enquote {\bibinfo {title} {{Skewed and
  double distributions in pion and nucleon}},}\ }\href {\doibase
  10.1103/PhysRevD.60.114017} {\bibfield  {journal} {\bibinfo  {journal} {Phys.
  Rev.}\ }\textbf {\bibinfo {volume} {D60}},\ \bibinfo {pages} {114017}
  (\bibinfo {year} {1999})},\ \Eprint {http://arxiv.org/abs/hep-ph/9902451}
  {arXiv:hep-ph/9902451 [hep-ph]} \BibitemShut {NoStop}%
%%CITATION = HEP-PH/9902451;%%
\bibitem [{\citenamefont {{A. V. Belitsky, D. M\"uller, A. Kirchner, and A.
  Sch\"afer}}(2001)}]{Belitsky:2000vk}%
  \BibitemOpen
  \bibfield  {author} {\bibinfo {author} {\bibnamefont {{A. V. Belitsky, D.
  M\"uller, A. Kirchner, and A. Sch\"afer}}},\ }\bibfield  {title} {\enquote
  {\bibinfo {title} {{Twist three analysis of photon electroproduction off
  pion}},}\ }\href {\doibase 10.1103/PhysRevD.64.116002} {\bibfield  {journal}
  {\bibinfo  {journal} {Phys. Rev.}\ }\textbf {\bibinfo {volume} {D64}},\
  \bibinfo {pages} {116002} (\bibinfo {year} {2001})},\ \Eprint
  {http://arxiv.org/abs/hep-ph/0011314} {arXiv:hep-ph/0011314 [hep-ph]}
  \BibitemShut {NoStop}%
%%CITATION = HEP-PH/0011314;%%
\bibitem [{\citenamefont {Tiburzi}(2004)}]{Tiburzi:2004qr}%
  \BibitemOpen
  \bibfield  {author} {\bibinfo {author} {\bibfnamefont {B.~C.}\ \bibnamefont
  {Tiburzi}},\ }\bibfield  {title} {\enquote {\bibinfo {title} {{Double
  distributions: Loose ends}},}\ }\href {\doibase 10.1103/PhysRevD.70.057504}
  {\bibfield  {journal} {\bibinfo  {journal} {Phys. Rev.}\ }\textbf {\bibinfo
  {volume} {D70}},\ \bibinfo {pages} {057504} (\bibinfo {year} {2004})},\
  \Eprint {http://arxiv.org/abs/hep-ph/0405211} {arXiv:hep-ph/0405211 [hep-ph]}
  \BibitemShut {NoStop}%
%%CITATION = HEP-PH/0405211;%%
\bibitem [{\citenamefont {Frederico}\ and\ \citenamefont
  {Miller}(1992)}]{Frederico:1992ye}%
  \BibitemOpen
  \bibfield  {author} {\bibinfo {author} {\bibfnamefont {T.}~\bibnamefont
  {Frederico}}\ and\ \bibinfo {author} {\bibfnamefont {G.~A.}\ \bibnamefont
  {Miller}},\ }\bibfield  {title} {\enquote {\bibinfo {title} {{Null plane
  phenomenology for the pion decay constant and radius}},}\ }\href {\doibase
  10.1103/PhysRevD.45.4207} {\bibfield  {journal} {\bibinfo  {journal} {Phys.
  Rev.}\ }\textbf {\bibinfo {volume} {D45}},\ \bibinfo {pages} {4207--4213}
  (\bibinfo {year} {1992})}\BibitemShut {NoStop}%
%%CITATION = PHRVA,D45,4207;%%
\bibitem [{\citenamefont {Tiburzi}\ and\ \citenamefont
  {Miller}(2003{\natexlab{a}})}]{Tiburzi:2002tq}%
  \BibitemOpen
  \bibfield  {author} {\bibinfo {author} {\bibfnamefont {B.~C.}\ \bibnamefont
  {Tiburzi}}\ and\ \bibinfo {author} {\bibfnamefont {G.~A.}\ \bibnamefont
  {Miller}},\ }\bibfield  {title} {\enquote {\bibinfo {title} {{Generalized
  parton distributions and double distributions for $q \overline{q}$ pions}},}\
  }\href {\doibase 10.1103/PhysRevD.67.113004} {\bibfield  {journal} {\bibinfo
  {journal} {Phys. Rev.}\ }\textbf {\bibinfo {volume} {D67}},\ \bibinfo {pages}
  {113004} (\bibinfo {year} {2003}{\natexlab{a}})},\ \Eprint
  {http://arxiv.org/abs/hep-ph/0212238} {arXiv:hep-ph/0212238 [hep-ph]}
  \BibitemShut {NoStop}%
%%CITATION = HEP-PH/0212238;%%
\bibitem [{\citenamefont {Tiburzi}\ \emph {et~al.}(2004)\citenamefont
  {Tiburzi}, \citenamefont {Detmold},\ and\ \citenamefont
  {Miller}}]{Tiburzi:2004mh}%
  \BibitemOpen
  \bibfield  {author} {\bibinfo {author} {\bibfnamefont {B.~C.}\ \bibnamefont
  {Tiburzi}}, \bibinfo {author} {\bibfnamefont {W.}~\bibnamefont {Detmold}}, \
  and\ \bibinfo {author} {\bibfnamefont {G.~A.}\ \bibnamefont {Miller}},\
  }\bibfield  {title} {\enquote {\bibinfo {title} {{Double distributions for
  the proton}},}\ }\href {\doibase 10.1103/PhysRevD.70.093008} {\bibfield
  {journal} {\bibinfo  {journal} {Phys. Rev.}\ }\textbf {\bibinfo {volume}
  {D70}},\ \bibinfo {pages} {093008} (\bibinfo {year} {2004})},\ \Eprint
  {http://arxiv.org/abs/hep-ph/0408365} {arXiv:hep-ph/0408365 [hep-ph]}
  \BibitemShut {NoStop}%
%%CITATION = HEP-PH/0408365;%%
\bibitem [{\citenamefont {Liu}\ and\ \citenamefont {Soper}(1993)}]{Liu:1992dg}%
  \BibitemOpen
  \bibfield  {author} {\bibinfo {author} {\bibfnamefont {H.-H.}\ \bibnamefont
  {Liu}}\ and\ \bibinfo {author} {\bibfnamefont {D.~E.}\ \bibnamefont
  {Soper}},\ }\bibfield  {title} {\enquote {\bibinfo {title} {{Implementation
  of the Leibbrandt-Mandelstam gauge prescription in the null plane bound state
  equation}},}\ }\href {\doibase 10.1103/PhysRevD.48.1841} {\bibfield
  {journal} {\bibinfo  {journal} {Phys. Rev.}\ }\textbf {\bibinfo {volume}
  {D48}},\ \bibinfo {pages} {1841--1851} (\bibinfo {year} {1993})}\BibitemShut
  {NoStop}%
%%CITATION = PHRVA,D48,1841;%%
\bibitem [{\citenamefont {Weinberg}(1966)}]{Weinberg:1966jm}%
  \BibitemOpen
  \bibfield  {author} {\bibinfo {author} {\bibfnamefont {S.}~\bibnamefont
  {Weinberg}},\ }\bibfield  {title} {\enquote {\bibinfo {title} {{Dynamics at
  infinite momentum}},}\ }\href {\doibase 10.1103/PhysRev.150.1313} {\bibfield
  {journal} {\bibinfo  {journal} {Phys. Rev.}\ }\textbf {\bibinfo {volume}
  {150}},\ \bibinfo {pages} {1313--1318} (\bibinfo {year} {1966})}\BibitemShut
  {NoStop}%
%%CITATION = PHRVA,150,1313;%%
\bibitem [{\citenamefont {Antonuccio}\ \emph {et~al.}(1997)\citenamefont
  {Antonuccio}, \citenamefont {Brodsky},\ and\ \citenamefont
  {Dalley}}]{Antonuccio:1997tw}%
  \BibitemOpen
  \bibfield  {author} {\bibinfo {author} {\bibfnamefont {F.}~\bibnamefont
  {Antonuccio}}, \bibinfo {author} {\bibfnamefont {S.~J.}\ \bibnamefont
  {Brodsky}}, \ and\ \bibinfo {author} {\bibfnamefont {S.}~\bibnamefont
  {Dalley}},\ }\bibfield  {title} {\enquote {\bibinfo {title} {{Light cone wave
  functions at small $x$}},}\ }\href {\doibase 10.1016/S0370-2693(97)01067-8}
  {\bibfield  {journal} {\bibinfo  {journal} {Phys. Lett.}\ }\textbf {\bibinfo
  {volume} {B412}},\ \bibinfo {pages} {104--110} (\bibinfo {year} {1997})},\
  \Eprint {http://arxiv.org/abs/hep-ph/9705413} {arXiv:hep-ph/9705413 [hep-ph]}
  \BibitemShut {NoStop}%
%%CITATION = HEP-PH/9705413;%%
\bibitem [{\citenamefont {Tiburzi}\ and\ \citenamefont
  {Miller}(2003{\natexlab{b}})}]{Tiburzi:2002sx}%
  \BibitemOpen
  \bibfield  {author} {\bibinfo {author} {\bibfnamefont {B.~C.}\ \bibnamefont
  {Tiburzi}}\ and\ \bibinfo {author} {\bibfnamefont {G.~A.}\ \bibnamefont
  {Miller}},\ }\bibfield  {title} {\enquote {\bibinfo {title} {{Current in the
  light front Bethe-Salpeter formalism. 2. Applications}},}\ }\href {\doibase
  10.1103/PhysRevD.67.054015} {\bibfield  {journal} {\bibinfo  {journal} {Phys.
  Rev.}\ }\textbf {\bibinfo {volume} {D67}},\ \bibinfo {pages} {054015}
  (\bibinfo {year} {2003}{\natexlab{b}})},\ \Eprint
  {http://arxiv.org/abs/hep-ph/0210305} {arXiv:hep-ph/0210305 [hep-ph]}
  \BibitemShut {NoStop}%
%%CITATION = HEP-PH/0210305;%%
\bibitem [{\citenamefont {Sawicki}(1991)}]{Sawicki:1991sr}%
  \BibitemOpen
  \bibfield  {author} {\bibinfo {author} {\bibfnamefont {M.}~\bibnamefont
  {Sawicki}},\ }\bibfield  {title} {\enquote {\bibinfo {title} {{Light front
  limit in a rest frame}},}\ }\href {\doibase 10.1103/PhysRevD.44.433}
  {\bibfield  {journal} {\bibinfo  {journal} {Phys. Rev.}\ }\textbf {\bibinfo
  {volume} {D44}},\ \bibinfo {pages} {433--440} (\bibinfo {year}
  {1991})}\BibitemShut {NoStop}%
%%CITATION = PHRVA,D44,433;%%
\bibitem [{\citenamefont {Bakker}\ \emph {et~al.}(2001)\citenamefont {Bakker},
  \citenamefont {Choi},\ and\ \citenamefont {Ji}}]{Bakker:2000pk}%
  \BibitemOpen
  \bibfield  {author} {\bibinfo {author} {\bibfnamefont {B.~L.~G.}\
  \bibnamefont {Bakker}}, \bibinfo {author} {\bibfnamefont {H.-M.}\
  \bibnamefont {Choi}}, \ and\ \bibinfo {author} {\bibfnamefont {C.-R.}\
  \bibnamefont {Ji}},\ }\bibfield  {title} {\enquote {\bibinfo {title}
  {{Regularizing the fermion loop divergencies in the light front meson
  currents}},}\ }\href {\doibase 10.1103/PhysRevD.63.074014} {\bibfield
  {journal} {\bibinfo  {journal} {Phys. Rev.}\ }\textbf {\bibinfo {volume}
  {D63}},\ \bibinfo {pages} {074014} (\bibinfo {year} {2001})},\ \Eprint
  {http://arxiv.org/abs/hep-ph/0008147} {arXiv:hep-ph/0008147 [hep-ph]}
  \BibitemShut {NoStop}%
%%CITATION = HEP-PH/0008147;%%
\bibitem [{\citenamefont {Lepage}\ and\ \citenamefont
  {Brodsky}(1980)}]{Lepage:1980fj}%
  \BibitemOpen
  \bibfield  {author} {\bibinfo {author} {\bibfnamefont {G.~P.}\ \bibnamefont
  {Lepage}}\ and\ \bibinfo {author} {\bibfnamefont {S.~J.}\ \bibnamefont
  {Brodsky}},\ }\bibfield  {title} {\enquote {\bibinfo {title} {{Exclusive
  processes in perturbative quantum chromodynamics}},}\ }\href {\doibase
  10.1103/PhysRevD.22.2157} {\bibfield  {journal} {\bibinfo  {journal} {Phys.
  Rev.}\ }\textbf {\bibinfo {volume} {D22}},\ \bibinfo {pages} {2157} (\bibinfo
  {year} {1980})}\BibitemShut {NoStop}%
%%CITATION = PHRVA,D22,2157;%%
\bibitem [{\citenamefont {Jaus}(1999)}]{Jaus:1999zv}%
  \BibitemOpen
  \bibfield  {author} {\bibinfo {author} {\bibfnamefont {W.}~\bibnamefont
  {Jaus}},\ }\bibfield  {title} {\enquote {\bibinfo {title} {{Covariant
  analysis of the light front quark model}},}\ }\href {\doibase
  10.1103/PhysRevD.60.054026} {\bibfield  {journal} {\bibinfo  {journal} {Phys.
  Rev.}\ }\textbf {\bibinfo {volume} {D60}},\ \bibinfo {pages} {054026}
  (\bibinfo {year} {1999})}\BibitemShut {NoStop}%
%%CITATION = PHRVA,D60,054026;%%
\end{thebibliography}%

\end{document}